\documentclass[aip,preprint, jmp,showpacs,showkeys]{revtex4-1}

\usepackage{graphicx}
\usepackage{amsmath}
\usepackage{amssymb}
\usepackage{dcolumn}
\usepackage{bm}
\usepackage{color,soul}
\usepackage{overpic}
\usepackage{subfigure}
\usepackage[scanall]{psfrag}

\def \be{\begin{equation}}
\def \ee{\end{equation}}

\def \UU{\mathbf{u}}

\newcommand{\bv}[1]{\mathbf{#1}}
\definecolor{lightgray}{gray}{0.85}
\sethlcolor{lightgray}
\usepackage{setspace}

\begin{document}



\title{Boundary elements method for microfluidic two-phase flows in shallow channels.} 

\preprint{}


\author{Mathias \surname{Nagel}}
  \email{mathias.nagel@epfl.ch}
\author{Fran\c{c}ois \surname{Gallaire}}%
\affiliation{ Laboratory of Fluid Mechanics and Instabilities - EPFL Lausanne, Switzerland
}%




\newcommand{\order}[1]{$\mathcal{O}(#1)$}
\newcommand{\del}[2]{\frac{\partial #1}{\partial #2}}
\newcommand{\ddel}[2]{\frac{\partial^2 #1}{\partial {#2}^2}}

\date{\today}

\begin{abstract}

In the following work we apply the boundary element method to two-phase flows in shallow microchannels, where one phase is dispersed and does not wet the channel walls. 
These kinds of flows are often encountered in microfluidic Lab-on-a-Chip devices and characterized by low Reynolds and low capillary numbers. 

Assuming that these channels are homogeneous in height and have a large aspect ratio, we use depth-averaged equations to describe these two-phase flows using the Brinkman equation, which constitutes a refinement of Darcy's law. 
These partial differential equations are discretized and solved numerically using the boundary element method, where a stabilization scheme is applied to the surface tension terms, allowing for a less restrictive time step at low capillary numbers. 
The convergence of the numerical algorithm is checked against a static analytical solution and on a dynamic test case. Finally the algorithm is applied to the non-linear development of the Saffman-Taylor instability and compared to experimental studies of droplet deformation in expanding flows.
\end{abstract}


\maketitle 

\section{Introduction}

Microhydrodynamics is a branch of fluid dynamics that deals with slow viscous flows at small length scales. 
In recent years the research field of microfluidics investigated the possibilities that microhydrodynamics offers to perform chemistry or biology on a micrometric scale. 
Such efforts have led to an increasing number of Lab-On-A-Chip applications in the last ten years~\cite{Whitesides}.
In this course droplet microfluidics has emerged~\cite{seemann12}, because it exploits the laminar flow in microchannels to precisely control and steer operation on droplets, which act as highly parallelizable reaction chambers.

Microfluidic length scales are in the order of tens to hundreds of micrometers. When microfluidic channels are filled with two immiscible liquids, for instance water and oil, the viscosities are in the order of $\mu \approx 10^{-3} Pa\,s$ and surface tension or interfacial tension in the order of $\gamma \approx 10^{-2} Pa\,m$, depending on the fluid mixture and surfactants. 
Due to the small length scale the flow resistance in these channels is high, which is one reason why flow rates usually range between a few $nl/min$ to hundreds of $\mu l/min$ with flow rates in the order of $mm/s$. 
The Reynolds number, Re$=\frac{\rho U L}{\mu}$, is small and therefore, it is often a reasonable approximation to discard the non-linear inertial terms and to consider Stokes flow, which is described in section~\ref{mathmodel}.

However, when considering two-phase flow even in the Stokes regime, the dynamics become non-linear due to the free interface between both liquids. 
The non-linearity stems from domains of different viscosity separated by a mobile interface under surface tension. 

Two competing effects dominate the dynamics; one comes from viscous shear and the other from surface tension. The capillary number expresses the balance between viscosity and surface tension: $\; Ca = \frac{\mu U}{\gamma} \;$, which is considered here to be between $10^{-5}$ and $10^{-1}$.

Throughout the article we consider shallow channels that lie in a common plane. 
Instead of trying to resolve the full three-dimensional problem, we solve a depth-averaged problem, which is two-dimensional. 
For shallow channels the velocity profile in the thin direction (z-axis) is assumed to be parabolic, a hypothesis that is also used to derive Darcy's law in two-dimensions (x-y plane).
Darcy's law states that the flow velocity $\mathbf{u} $ is given by the pressure gradient divided by viscosity $\mu$ and a permeability coefficient $k^2$, $\nabla p = -\mu \mathbf{u}  k^2$.

Although there have been propositions to account for tangential surface stresses in Darcy's law~\cite{Borhan}, the inability to impose tangential stresses and velocities renders this approach incomplete.

In this work we propose the use of the Brinkman equations instead, which include a correction to the Darcy's law in form of the depth-averaged in-plane Laplacian, a reminiscence of the $2D$~Stokes equation.
For droplet flows the Brinkman equation was to our knowledge first proposed by Boos \textit{et al.}~\cite{BoosThess} and Bush~\cite{bush97}, who treated the flow induced by a thermo-capillary effect.

The Brinkman equation is solved with a boundary element method (BEM), which eliminates one more dimension turning the problem from a $2D$~differential equation into an integral equation on a $1D$~line. 
While BEM approaches have been followed for $3D$~Stokes flows~\cite{Bazhlekov}, for $2D$~Stokes flow~\cite{Wrobel} and Darcy~flow~\cite{Nadim}, recall that simulations of $2D$~Stokes flow cannot account for the confinement in the z-direction whereas Darcy's law becomes invalid close to boundaries and interfaces.
The use of the Brinkman equation requires high aspect ratios to justify depth-averaging but we will see that it might still accurately captures the dynamics for aspect ratios approaching $1$. Close to boundaries or interfaces the Brinkman equation gives much better results than Darcy's law, because it captures depth averaged boundary layers even if the averaged equations become inconsistent \cite{BoosThess}.

The derivation of the depth-averaged problem is presented in section~\ref{mathmodel} and the numerical method is described in section~\ref{numerical} together with a stabilization scheme for the surface tension on the interface and an acceleration using Gauss block pre-condensation and multi-core parallelism. 

The method is applied to the non-linear development of the Saffman-Taylor instability of finger formation and to the numerical modeling of two recent experimental studies of droplet deformation in section~\ref{results}. 
Section~\ref{conclusion} concludes with a brief discussion of the method and its results.

\section{Governing equations}
\label{mathmodel}

Throughout the article vectors and tensors are written in bold face unless they are represented by a Greek character. 
Scalars or components of vectors and matrices are written in normal face. 
All field variables are non-dimensionalized, using a characteristic length scale $L$, the pressure scale $P=\gamma_\textrm{ref}/L$ and the velocity scale $ U = \gamma_\textrm{ref} /\mu_c$, which are build using the continuous fluids viscosity $\mu_c$ and surface tension $\gamma_\textrm{ref}$.

Low Reynolds number flows are described by the $3D$~Stokes and continuity equation, where non-dimensional operators and variables in $3D$ are denoted with a tilde.

\be
	\lambda_\phi \Tilde \Delta \mathbf{\tilde u} -\nabla \tilde p = 0 \quad \textrm{and} \quad \tilde \nabla \cdot \mathbf{\tilde u}= 0.
	\label{eq:stokes3d}
\ee
The non-dimensional parameter $\lambda_\phi$ compares the viscosity of the considered fluid phase $\phi$ against the viscosity of the carrier fluid, $ \lambda_\phi = \mu_\phi / \mu_c $. 
For the dispersed phase $\phi=d$, $\lambda_d = \lambda=\mu_d / \mu_c $, and for the continuous phase $\phi=c$, $\lambda_c = 1$. 
Because of the small size and the horizontal alignment gravitational effects are neglected. 

\subsection{Brinkman model for depth-averaged flow}
The non-dimensional height is $h= H/L$ and is considered to be small, $h\ll1$.
As the flow is confined between two plates at a distance $h$, one considers only fluid motion in the $x-y$ plane and neglects the vertical velocity component, which is equivalent to the assumption of constant pressure in the $z$-direction.

Under this assumption the flow field writes $\bv{\tilde u}(x,y,z) = (u_x(x,y) f(z) ,\; u_y(x,y) f(z), \; 0)^T$. 
The two-dimensional velocity vector $\bv{u}(x,y)=\binom{u_x(x,y)}{u_y(x,y)}$ represents mean velocities, which demands $\int_0^h f(z) dz = h$.
With these assumptions eq.(\ref{eq:stokes3d}) can be written in terms of two-dimensional variables and operators:
\be
 			\lambda_\phi \left( \Delta \mathbf{u} + \bv{u} \ddel{f(z) }{z} \right) - \nabla p = 0 \quad \textrm{and} \quad \nabla \cdot \bv{u} = 0.
 			\label{eq:precursor}
\ee
When $h \ll 1$ the profile $f(z)$ becomes a parabolic Poiseuille profile and its second derivative in eq.(\ref{eq:precursor}) is known. 
Using the parabolic profile $f(z) = 6\, \frac{z}{h} \left(1-\frac{z}{h}\right)$ in eq.(\ref{eq:precursor}) and depth-averaging over $z$ we get the amalgam equation of Darcy equation and $2D$~Stokes equation given in equation (\ref{fct:brinkman}), which is called Brinkman equation and was first applied in granular media flows~\cite{brinkman},
\be
	\lambda_\phi( \Delta \bv{u} -k^2 \bv{u}) - \nabla p = 0, \quad \nabla \cdot \bv{u} = 0,	\quad k= \frac{\sqrt{12}}{h}.
	\label{fct:brinkman}
\ee 

\begin{figure}[htb]
	\centering
	\includegraphics[width=\textwidth]{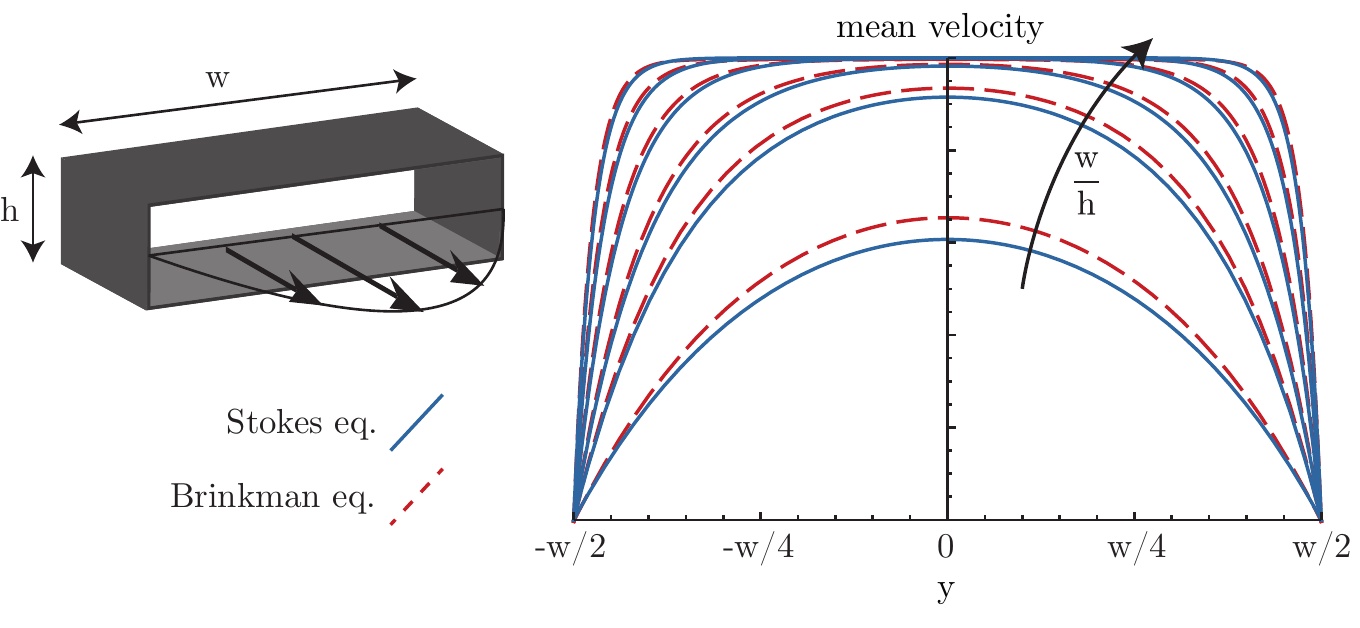}
	\caption{Depth-averaged velocity profiles across a parallel channel of width $W$ for $\partial p/\partial x = -12/h^2$. The full line represents the solution of the $3D$~Stokes equation and the dashed line the solution of the Brinkman equation. The aspect ratio of the profiles shown from bottom up, $w/h = 1,2,3,5,8,12$.}
	\label{fig:3dchannelcompare}
\end{figure}
We shall briefly illustrate the advantage of the Brinkman equation through a comparison of its solution for a flow in a rectangular duct of width $w$ with the $3D$~Stokes solution. The exact solution can be found by separation of variables and is given for instance in Langlois and Deville \cite{deville14}.
Depth-averaging the solution of the $3D$~Stokes equation gives the mean velocity across the channel.
\begin{equation}
	< \tilde{u} >\; = -\frac{\partial p}{\partial x} \frac{h^2}{12} \left( 1 - \frac{96}{\pi^4} \sum \frac{\cosh \left( (1 + 2 n) \pi y/h \right)}{(1+2 n)^4 \cosh\left((1 + 2 n) \pi w/(2 h) \right)}  \right).
	\label{eq:pois3d}
\end{equation}
In comparison, the mean velocity using of the depth-averaged Brinkman equation is:
\begin{equation}
	\centering
	u = - \frac{\partial p}{\partial x} \frac{h^2}{12} \left( 1 - \frac{\cosh(\sqrt{12} y/h)}{\cosh(\sqrt{12}w/(2h))} \right).
	\label{eq:pois2d}
\end{equation}
Both solutions show at leading order a hyperbolic cosine with similar prefactors, $96/\pi^4 \approx 0.986$ for the Stokes equation instead of $1$ for the Brinkman equation and in the hyperbolic cosine a factor $\pi \approx 3.14$ instead of  $\sqrt{12} \approx 3.46$. The depth-averaged velocity profiles are plotted in figure \ref{fig:3dchannelcompare} for different aspect ratios. One observes that the solution from the Brinkman equation tends the solution of the $3D$~Stokes equation as the aspect ratio increases. Already for square channels, $w/h=1$, the solutions are not too far from each other. Whereas a comparison with the Darcy equation, which is constant in $y$, gives $\overline{u}_{Darcy} = - \frac{\partial p}{\partial x} \frac{h^2}{12}$.
Far away from the walls the Darcy equation gives correct results for high aspect ratios but fails near walls and for a moderate confinement.

In a more detailed analysis Gallaire \textit{et al.}~\cite{Gallaire} showed that even in the complex thermo-capillary flow around a droplet the averaged model agrees almost perfectly with $3D$~Stokes. 
Including the in-plane Laplacian yields two important improvements in comparison to Darcy's law: 1) Tangential velocities and stress can be imposed on boundaries and 2) there appears a boundary layer near walls and interfaces that scales like $h$, the non-dimensional height of the channel. 

\subsection{In-flow and out-flow boundary conditions}

Boundary conditions of the single-phase problem prescribe either the stress or the velocity. 
The typical no-slip boundary condition on channel walls is $\bf{u}=0$. In contrast to Darcy flow, the Brinkman model imposes normal and tangential velocities. 
The normal and tangent are given by a vector that contains their projections on the $x$ and $y$ axis, e.g. $\mathbf{n} = (n_x, n_y)^T$. 

As typical inflow boundary condition the solution of the Brinkman equation in a straight channel flow is used.
For a straight inflow boundary of length $w$ parameterized by $s$, whose origin is in the middle of the boundary:
\be
	u_{in}(s) = Ca \frac{\cosh(k \, w/2) - \cosh(k \, s)}{\cosh(k \, w/2) - 1}.
\ee
It is worth observing that the dimensionless inflow velocity is represented by the capillary number $Ca$ because the velocity is non-dimensionalized by $U = \gamma_{\textrm{ref}}/\mu_c$. 
Integration of the velocity field along the inflow boundary relates the dimensional flow rate $Q$ to the inflow capillary number. 
\be
	Ca =  \frac{ Q \, \mu_c }{L^2\, w\,h\,  \gamma } \frac{\cosh(k\, w/2) -1}{\cosh(k\, w/2)-\sinh(k\, w/2)/(k \,w/2)}.
\ee
The outflow boundary of parallel channel flow with constant pressure is imposed by zero tangential velocity, $\mathbf{u \cdot t} = 0$ and constant normal stress $\mathbf{n} \cdot \sigma \cdot \mathbf{n} = p_0$. 

\subsection{Droplet interface condition}
\label{sec:wt_films}
Droplet interface conditions are continuity of velocity and discontinuity of interface stresses. 
The surface or interface stresses shall also be denoted by $\mathbf{f} = \sigma \bv{n}$, whose components are surface stresses $f_x, f_y$.
Over the interface the normal stress is discontinuous due to surface tension and curvature.
The tangential stress can also be discontinuous due to a varying surface tension. The varying surface tension is introduced by $\gamma = \tilde \gamma(\mathbf{x})/\tilde \gamma_{\textrm{ref}}$.
The two principal curvatures are the in-plane curvature $\kappa_{||}$ and the meniscus curvature in the thin direction $\kappa_\perp$. 
\be
	[\![ \sigma \bv{n}  ]\!] = \mathbf{[\![ f ]\!]} = \gamma \left(  \kappa_{||} \frac{\pi}{4}+\kappa_\perp \right) \bv n +\frac{d \gamma}{d s}\bv t  = \frac{ d (\gamma \bf{t})}{ d s } -\gamma \left( 1-\frac{\pi}{4} \right) \frac{d \mathbf{t}}{d s} + \gamma \kappa_\perp \mathbf{n}.
	\label{eq:laplace}
\ee
In accordance with real droplet microfluidic systems we assume a non-wetting dispersed phase, thus the out-of-plane meniscus is approximately a half-circle, $\kappa_\perp = \frac{2}{h}$, and the in-plane curvature $\kappa_{||}$ is corrected by a $\frac{\pi}{4}$ term, which was derived by Park and Homsy~\cite{park84}.
With a perspective on discretization the stress jump is also expressed in terms of Frenet equations, which is proposed by Tryggvason and co-authors~\cite{zaleski}. 

In this formulation of the interface stresses jump we neglect the effect of dynamic film formation, which changes the out-of-plane curvature $\kappa_\perp$. This effect, which depends in a non-linear way on the capillary number, see Park \textit{et al.} will be discussed in the conclusion. 

\section{Numerical method}
\label{numerical}

In order to solve the two-phase flow numerically the domain has to be discretized. 
It is desirable to apply an interface-tracking scheme, as the interface represents a localized force. 
This force is relatively high because it competes with viscous forces that scale like the capillary number, which is small.

Using a diffusive interface instead of a discrete interface can lead to problems at low capillary number. For instance Carlson \textit{et al.}~\cite{amberg10} used a phase field method with a diffuse interface to study the dynamics of droplets in a bifurcation in $3D$ at low Reynolds numbers. They observed a dependence of the solution on the thickness of the interface and had to use a very small time step.
In order to avoid these difficulties we discretize the interface with mesh elements for the sake of precision and stability of the numerical algorithm. 

A Boundary Element Method is implemented, where only boundaries are discretized. As a consequence it is unnecessary to remesh the whole domain as the interface evolves.
For a complete description of the method and applications in viscous flow one may consult the book by Pozrikidis~\cite{pozrikidis}.
In this book the Brinkman equation is mentioned as a mean to compute unsteady flows, where a discretized time derivative due to the local acceleration is represented by $k^2 \mathbf{u}$. In the following section we elaborate the two-phase flow boundary integral form of the Brinkman equation and its boundary conditions. The procedure closely follows that for $2D$~Stokes flow \cite{wrobel} but we include it for a complete and comprehensive presentation of the method.

\subsection{Boundary integral formulation}
Transformation of the Brinkman equation into a boundary integral form is done by integration by parts, which requires expressing the Brinkman equation from eq.(\ref{fct:brinkman}) using the divergence of the in-plane stress tensor.

The in-plane stress tensor $\sigma$ is,
\begin{equation}
	\sigma = \left( \begin{array}{cc}
	2\lambda_\phi \frac{\partial u_x}{\partial x} -p & \lambda_\phi \left( \frac{\partial u_x}{\partial y} + \frac{\partial u_y}{\partial x} \right) \\   \lambda_\phi \left( \frac{\partial u_x}{\partial y} + \frac{\partial u_y}{\partial x} \right) & 2\lambda_\phi \frac{\partial u_y}{\partial y} -p
	\end{array} \right).
\end{equation}
Hence the Brinkman equation can be expressed as:
\begin{equation}
		\nabla \cdot \sigma -k^2 \bv{u} = 0, \quad \nabla \cdot \bv{u} = 0.
	\label{fct:brinkman2}
\end{equation}
As a first step we integrate the Brinkman eq.(\ref{fct:brinkman2}) and continuity equation over a vector field $\mathbf{v}$ and a scalar field $q$ on a domain $\Omega$ one obtains:
\be
	\int \limits_\Omega \left( \Bigl( \nabla \cdot \sigma -k^2 \bv{u} \Bigr) \cdot \mathbf{v}+ \lambda_\phi \, q \nabla \cdot \mathbf{u}  \right) dA = 0.
	\label{eq:firsttest}
\ee
Here $\mathbf{v}$ and $q$ can be seen as test functions, as in the Finite Elements Method. The continuity equation has been multiplied by $\lambda_\phi$, which will be necessary later on for the droplet interface boundary condition. If the velocities $\mathbf{u}$ and pressure $p$ fulfill this equation for whatever choice of test function they also solve the initial equation system eq.(\ref{fct:brinkman}).

Performing integration by parts on eq.(\ref{eq:firsttest}) gives,
\begin{eqnarray}
	\nonumber
	&&\int \limits_\Omega \Bigl( \lambda_\phi \left( -\nabla \UU : \nabla \mathbf{v} -k^2 \UU \cdot \mathbf{v} \right) + p \, \nabla \cdot \mathbf{v} + \lambda_\phi\, q \nabla \cdot \UU \Bigr) dA  \\
	&+&\oint \limits_\omega \left( \sigma \mathbf{n} \cdot \mathbf{v} \right) ds = 0. 
	\label{eq:partint1}
\end{eqnarray}
The term that appears as a boundary integral on the boundary $\omega$ of the domain $\Omega$ appears as a surface stresses $\sigma \mathbf{n}$ or $\mathbf{f}$. Performing integration by parts once more makes $\UU$ and $\mathbf{v}$ exchange roles. 
\begin{eqnarray}
	\nonumber
	&&\int \limits_\Omega \Bigl( \lambda_\phi \left( \Delta \mathbf{v} -k^2 \mathbf{v} -\nabla q \right) \cdot \UU + \nabla \cdot \mathbf{v} \; p \Bigr) dA \\
	 &+&\oint \limits_\omega \left( \sigma \mathbf{n} \cdot \mathbf{v} - \lambda_\phi \left( \begin{array}{cc}
	2\frac{\partial v_x}{\partial x} -q & \frac{\partial v_x}{\partial y} + \frac{\partial v_y}{\partial x}  \\ \frac{\partial v_x}{\partial y} + \frac{\partial v_y}{\partial x} & 2\frac{\partial v_y}{\partial y} -q
	\end{array} \right) \mathbf{n} \cdot \UU \right) ds = 0.
	\label{eq:secondtest}
\end{eqnarray}
After integrating by parts twice the Laplace operator that acted on the velocity $\UU$ has been transferred to $\mathbf{v}$. Since the vector field $\mathbf{v}$ takes the role of a velocity in a reappearing Brinkman equation, $q$ can be interpreted as a pressure and the new boundary integral term can be interpreted as a surface stress $\tau \mathbf{n}$:
\begin{equation}
	\tau \mathbf{n} = \left( \begin{array}{cc}
	2\frac{\partial v_x}{\partial x} -q & \frac{\partial v_x}{\partial y} + \frac{\partial v_y}{\partial x}  \\ \frac{\partial v_x}{\partial y} + \frac{\partial v_y}{\partial x} & 2\frac{\partial v_y}{\partial y} -q
	\end{array} \right) \mathbf{n}.
\end{equation}
When $\mathbf{v}$, $q$ and $\tau$ solve the Brinkman equation the domain integral disappears and only boundary integrals are left. 
Green's functions of the Brinkman equation is used a test function because they have a Dirac like forcing, which ensure that in a numerical discretization of the problem the highest elements are usually the diagonal elements of the matrix, leading to a low condition number of the problem.
A Dirac forcing on the boundary also allows expressing the domain integral solely based on the velocities at the location of the Dirac distribution.
 
\subsection{Green's functions}
In order to solve for the unknown velocities and stresses two Green's functions are necessary. A third Green's function can be used to compute the pressure in the domain.
As a consequence of the Dirac forcing $\delta(\mathbf{x}_0)$ the velocity, pressure and stress fields have singularities at $\mathbf{x}_0$, where their respective values go to infinity.  

In the following we present at first the two Green's function of the Brinkman equation with a Dirac forcing of the stress equation in $x$ or $y$ direction.
Taking two velocity vectors $\mathbf{G}_a$, two pressures $P_a$ and two stress tensor fields $\mathbf{T}_a$, with $a=1,2$, and which verify: 
\be
	\nabla \cdot \mathbf{T}_a - k^2 \mathbf{G}_a = \Delta \mathbf{G}_a -\nabla P_a - k^2 \mathbf{G}_a  = \mathbf{w}_a \, \delta( \mathbf{x}_0) \quad \textrm{and} \quad \nabla \cdot \mathbf{G}_a = 0.
	\label{eq:testeq}
\ee 
The Dirac forcing occurs at the position $\mathbf{x}_0$ with a vectorial weight $\mathbf{w}_a$. 
We are interested in two linear independent weight vectors $\mathbf{w}_1 = \binom{1}{0}$ and $\mathbf{w}_2 = \binom{0}{1}$.

The solutions are given in indexed form, first index $a$ for the corresponding weight function $\mathbf{w}_a$ and $b,c$ for the $b^{th}$ component of a vector or the $(b,c)$ component in a matrix. 
The Green's function velocity field, pressure and stress tensor are taken from Pozrikidis \cite{pozrikidis}.
\be
	G_{ab} = \frac{1}{4 \pi} \left( A_1 \delta_{ab} - A_2 \frac{x_a\,x_b}{r^2} \right).
	\label{eq:fundvel}
\ee
Introducing $A_1$ and $A_2$, which are functions of aspect ratio times distance $k\,r$ and are expressed by modified Bessel functions of second kind $K_0(k\, r)$ and $K_1(k\,r)$. In the above and following definitions $\delta_{ab}$ is the Kronecker delta and not to be confused the Dirac distribution $\delta(\mathbf{x}_0)$.
\be
	A_1 = 2\Bigl( \frac{1}{k^2 r^2} -\frac{K_1(k\,r)}{k\,r} - K_0(k\,r) \Bigr),\; A_2 = 2\Bigl( \frac{2}{k^2 r^2} -2\frac{K_1(k\,r)}{k\,r} - K_0(k\,r)  \Bigr)
	\label{eq:a1a2}
\ee
The function $A_1(k r)$ behaves like a $\log(k\,r)$ close to $r\approx 0$ and is therefore weakly singular.
Using this result in eq.(\ref{eq:testeq}) leads to the associated pressure field $q$. 
\be
	P_a = -\frac{x_a}{2\pi r^2}.
\ee
From velocity and pressure fields one derives the associated stress tensor, \linebreak $T_{abc} = -P_a \delta_{bc} + \frac{\partial \mathbf{G}_{ab}}{\partial x_c}$:
\be
	T_{abc} = \frac{\delta_{bc} x_a}{2 \pi r^2} (1-A_2) + \frac{\delta_{ab} x_c+ \delta_{ac} x_b}{2 \pi r^2}(K_1(k\,r) \, k\,r-A_2)-\frac{x_a x_b x_c}{\pi r^4}(K_1(k\,r) \, k\,r-2A_2).
	\label{eq:teststress}
\ee
Inserting now $\mathbf{v} = \mathbf{G}_a$ and $\tau = \mathbf{T}_a$ with $a=1$ or $2$ into eq.(\ref{eq:secondtest}) gives two independent boundary integral equations with a forcing either in the $x$ or $y-$direction.
Because of the Dirac forcing the domain integral does not disappear completely but is confined to a single point and eq.(\ref{eq:secondtest}) becomes:
\be
	\lambda_\phi \int \limits_\Omega \Bigl( \delta(\mathbf{x}_0) \mathbf{w}_a \cdot \UU \Bigr) dA = S \mathbf{w}_a \cdot \UU(\mathbf{x}_0) = \oint \limits_\omega \Bigl( \lambda_\phi \mathbf{T}_a \mathbf{n} \cdot \UU - \sigma \mathbf{n} \cdot \mathbf{G}_a \Bigr) ds.
	\label{eq:bie1}
\ee
The domain integral over the Dirac selects the value of $\mathbf{u}$ at $\mathbf{x}_0$, the position where the Green's functions was forced. 
Depending on whether $\mathbf{x}_0$ is in the domain, on the boundary or outside the prefactor $S$ is $\lambda_\phi, \frac{\lambda_\phi}{2}$ or $0$. 
\vspace*{0.5cm}
When it is desired to determine the pressure in the domain, one uses a third fundamental solution with a Dirac distribution forcing of the continuity equation, which corresponds to a point source in the continuity equation at $\mathbf{x}_0$. Instead of eq.(\ref{eq:testeq}) the Green's function solves:
\be
	\nabla \cdot \mathbf{T}_3 - k^2 \mathbf{G}_3 = \Delta \mathbf{G}_3 -\nabla P_3 - k^2 \mathbf{G}_3  = 0 \quad \textrm{and} \quad \nabla \cdot \mathbf{G}_3 = \delta( \mathbf{x}_0).
	\label{eq:testeq3}
\ee
	This point source solution is also solution of the Darcy equation and it's fields are:
\be
	P_3 = - \frac{k^2}{4\pi} \log(r^2), \; G_{3b} = \frac{x_j}{2 \pi r^2}, \; T_{3bc} = \frac{1}{4 \pi}\left( \left( k^2 \log(r^2)+\frac{4}{r^2} \right) \delta_{bc}-8\frac{x_b x_c}{r^4} \right). 
\ee
Inserted in eq.(\ref{eq:secondtest}) this gives:
\be
	\int \limits_\Omega \delta(\mathbf{x}_0) \, p \, dA = S \, p(\mathbf{x}_0) = \oint \limits_\omega \Bigl( \lambda_\phi \mathbf{T}_3 \mathbf{n} \cdot \UU -\sigma \mathbf{n} \cdot \mathbf{G}_3 \Bigr) ds.
	\label{eq:thirdtest}
\ee
Again the prefactor $S$ depends whether $\mathbf{x}_0$ is in the domain, on the boundary or outside, $S=(1, \frac{1}{2}, 0)$. Evaluating this integral equation allows to calculate the pressure at a given point in the domain once the stresses and velocities at the boundary are known.

\subsection{Boundary integral for two-phase flow}
\begin{figure}[htb]
\begin{center}
	\includegraphics[scale=1.0]{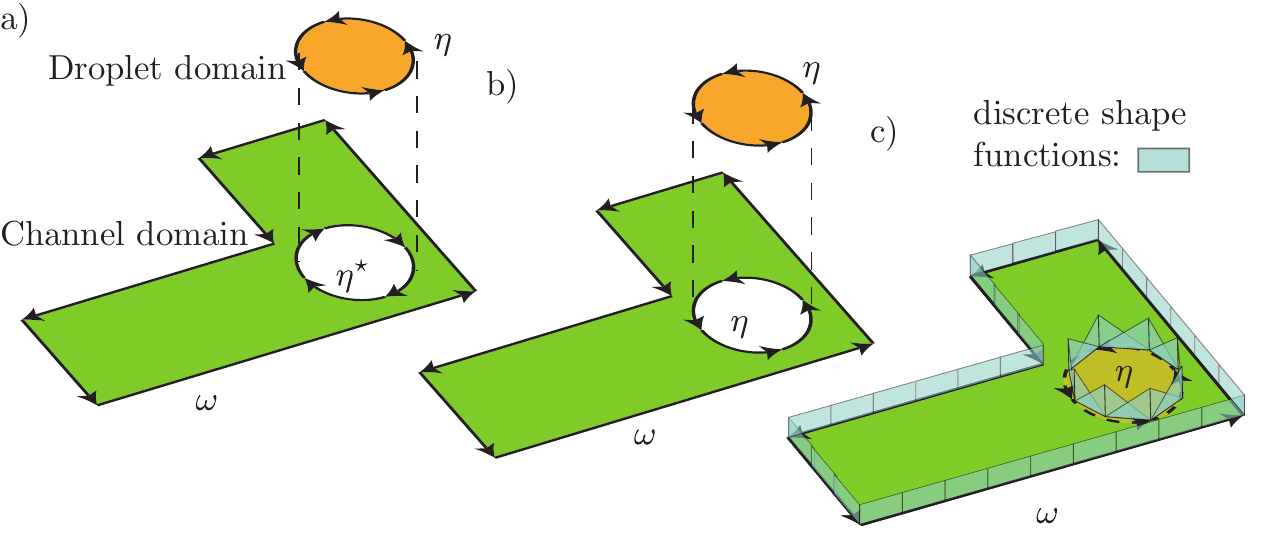}
\caption{Schematic representation of how the two fluid domains are coupled. Arrows depict the boundary integral paths. a) Two separate domains. b) Changing the orientation of the channels internal boundary, $\eta \rightarrow \eta^\star$. c) Combination of both integral contributions on the interface using the interface conditions in eq.(\ref{eq:laplace}). Piece-wise constant and piece-wise linear shape functions are sketched.}
\label{fig:domoint}
\end{center}
\end{figure}

When dealing with two-phase flows the two fluid domains need to be coupled by their common boundaries. 
The channels internal boundary is the same as the droplets boundary, except for the sense of orientation. 
The orientation determines the direction of the normal and therefore changing the sign of one of the normals makes both integral paths identical. 
One includes the droplet into the boundary integral formulation by summing the contributions of the boundary integral of the droplet $\eta$ and the channel domain, where the channel domains internal boundary $\eta^\star$ has a reversed integration path, i.e. inverted normals. Boundary integration is performed counter-clockwise around a domain, illustrated in Figure \ref{fig:domoint}.

\be
	\oint \limits_\omega \Bigl( \mathbf{T}_a \mathbf{n} \cdot \mathbf{u} - \sigma \mathbf{n} \cdot \mathbf{G}_a \Bigr) ds + \oint \limits_\eta \Bigl( -\mathbf{T}_a \mathbf{n} \cdot \mathbf{u} + \sigma \mathbf{n} \cdot \mathbf{G}_a \Bigr) ds = \frac{1}{2}\mathbf{w}_a \cdot \mathbf{u}(\mathbf{x}_0).
	\label{eq:bem_continous}
\ee

In the channel domain the viscosity ratio is $\lambda_c = 1$ and in the droplet $\lambda_d=\lambda$. The dispersed phase boundary integral problem is:

\be
	\oint \limits_\eta \Bigl( \lambda \mathbf{T}_a \mathbf{n} \cdot \mathbf{u}- \sigma \mathbf{n} \cdot \mathbf{G}_a \Bigr) ds = \frac{\lambda}{2} \mathbf{w}_a \cdot \mathbf{u}(\mathbf{x}_0).
	\label{eq:bem_dispersed}
\ee

Combining eq.(\ref{eq:bem_continous}) and (\ref{eq:bem_dispersed}) by adding the contributions under the integral along $\eta$ gives,

\be
	\oint \limits_\omega \Bigl( \mathbf{T}_{a} \mathbf{n} \cdot \UU -\sigma \bv{n} \cdot \mathbf{G}_a \Bigr) ds + \oint \limits_\eta \Bigl( \mathbf{T}_a \mathbf{n} \cdot (\lambda -1) \UU - [\![ \sigma \bv{n} ]\!] \cdot \mathbf{G}_a \Bigr) ds = S \mathbf{w}_a\cdot \mathbf{u}(\mathbf{x}_0).
	\label{bem:wtdrop}
\ee

If the singularity at $\mathbf{x}_0$ is located on the outer boundary the coefficient $S = \frac{1}{2}$ and if the singularity is located on the interface $S = \frac{1+\lambda}{2}$. 
Prescribing the jump in viscosity $(\lambda-1)$ and jump in surface stresses $[\![ \sigma \bv{n} ]\!]$, see eq.(\ref{eq:laplace}), specifies the fluid-fluid interface conditions and poses the combined problem.

A significant simplification is obtained for $\lambda=1$, where from eq.(\ref{bem:wtdrop}) only a diagonal matrix and right hand side remains,
\be
	\oint \limits_\eta \Bigl( [\![ \sigma \bv{n} ]\!] \cdot \mathbf{G}_a \Bigr) ds = -\mathbf{w}_a \cdot \mathbf{u}(\mathbf{x}_0).
\ee 
In the absence of walls the velocity at the point $\mathbf{x}_0$ depends only on known quantities and does not require the solution of a matrix. Even in the presence of outer boundaries the evolution problem can be reduced by one matrix inversion in a problem that involves only matrix vector multiplications, which will be shown in section \ref{sec:blockelim}.
We do not restrict our method to $\lambda=1$ it is worth mentioning that Zhu \textit{et al.}~\cite{Power} have used this in the case of $3D$ droplet motion. 

\subsection{Spatial discretization}
\label{stabilize}
The boundaries are discretized by piece-wise straight line elements with local shape functions whose amplitudes are to be determined.
Fixed boundary conditions like walls and inflow and outflow boundaries are discretized with piece-wise constant shape functions, whereas the droplet is discretized by piece-wise linear shape functions.
The boundary integral form of the Brinkman equation, eq.(\ref{bem:wtdrop}) is discretized with a collocation method and integrated by Gauss-Legendre quadrature with $6$ nodes.

The quadrature uses $6$ weights $w_j$, which are evaluated at $6$ base points $s_j$. For instance applied to a generic real valued function $f(x)$ integrated between $x_1$ and $x_2$
\begin{eqnarray}
	& \int \limits_{x_1}^{x_2} f(x) dx \approx \frac{|x_2-x_1|}{2} \sum \limits_{j=1}^6 w_j f\left( \frac{x_2+x_1}{2} + s_j \frac{x_2-x_1}{2} \right),& \\
	\nonumber
	s  = & \big[\pm 0.932469514203152, \pm 0.661209386466265, \pm 0.238619186083197 \big],&\\
		\nonumber
	w =& \big[0.171324492379170, 0.360761573048139, 0.467913934572691 \big].&
\end{eqnarray}

Discretized normal and tangential vector $\mathbf{n}^i, \mathbf{t}^i$, element length $|\Delta x_i|$ and local variable $\chi^i_j$ are given as:
\begin{equation}
	\mathbf{t}^i = \frac{\mathbf{x}_{i+1}-\mathbf{x}_i}{|\mathbf{x}_{i+1}-\mathbf{x}_i|}, \; \mathbf{n}^i = \left( \begin{array}{c} -t^i_y \\ t^i_x \end{array} \right),\; |\Delta x_i| = |\mathbf{x}_{i+1}-\mathbf{x}_i|, \; \chi^i_j = \frac{\mathbf{x}_{i+1}+\mathbf{x}_i}{2} +s_j \frac{\mathbf{x}_{i+1}-\mathbf{x}_i}{2}.
\end{equation}

The integral equation, eq.(\ref{bem:wtdrop}), turns into a double summation, the sum of the quadrature points, summed over all $M$ fixed elements and $N$ droplet elements.
\begin{eqnarray}
	\nonumber
S_l \mathbf{w}_a \cdot \mathbf{u}^l &=& \sum \limits_{i=1}^M \sum \limits_{j=1}^6 \frac{w_j |\Delta x_i|}{2} \left( \mathbf{T}_a^l(\chi_j^i-\mathbf{x}_l) \mathbf{n}^i \cdot \mathbf{u}^i - \mathbf{f^i} \cdot \mathbf{G}_a^l(\chi_j^i-\mathbf{x}_l) \right)\\
\nonumber
& + & \sum \limits_{i=M+1}^{M+N} \sum \limits_{j=1}^6 \frac{w_j |\Delta x_i|}{2} \bigg( \mathbf{T}_a^l(\chi_j^i-\mathbf{x}_l) \mathbf{n}^i \cdot (\lambda-1)\left( \frac{1-s_j}{2} \mathbf{u}^i + \frac{1+s_j}{2} \mathbf{u}^{i+1}\right) \\
&-&  \left( \frac{1-s_j}{2} [\![ \mathbf{f}^i  ]\!] + \frac{1+s_j}{2} [\![ \mathbf{f}^{i+1}  ]\!] \right) \cdot \mathbf{G}_a^l(\chi_j^i-\mathbf{x}_l) \bigg).
\label{eq:bem_discrete}
\end{eqnarray}

In eq.(\ref{eq:bem_discrete}), there are $2 \times (M+N)$ unknowns to be solved for; $2\times M$ velocities or surface stresses on the channel boundaries and $2\times N$ velocities on the drop interface.
For each position $l$ of the test functions on one of the $M+N$ nodes, there are two discretized equations, eq.(\ref{eq:bem_discrete}), for $a=1$ or $2$.
Summing up all known variables in the right hand side and forming a dense $2(M+N)\times 2(M+N)$ matrix with a vector of $2(M+N)$ unknowns, yields a linear equation system.

Due to the weak and strong singularities of the Green's functions, these functions diverge on the collocating element although the integral does not.
In numerically integration the singularities need special attention.
The test velocity $G_{aa}$ is weakly singular like a logarithm; in fact its asymptotic development in terms of Bessel functions contains a $\log(r)$. 
If the integration interval contains a singularity, the $\log(r)$ contribution is not evaluated in Gauss quadrature but integrated analytically.

The last member of eq.(\ref{eq:teststress}) for the stress $\mathbf{T}$ contains a $1/r$ singularity but its divergent behavior is perpendicular to the normal $\mathbf n$ and vanishes. To show this we consider an element collocating with a singularity, whose normals can be written as: $n_x = y/r, \; n_y = -x/r$.
\[
	\mathbf{T}_a \mathbf{n} = \frac{x_a}{\pi r^4} \left( \begin{array}{cc}
	x^2\; & xy \\ yx\; & y^2
\end{array} \right) \left( \begin{array}{c}
\frac{y}{r} \\ -\frac{x}{r}
\end{array} \right) +  \mathcal{O}(k^2 r^2) \approx  \frac{x_a}{\pi r^5}\left( \begin{array}{c}
	x^2 y - x^2 y\\ x y^2  - x y^2
\end{array} \right)= 0.
\]

Hence the strongly singular contribution cancels out, which allows setting them to zero on a collocating element and leaves an expression that can be numerically integrated. The numerical approximations of the Bessel functions with a subtractable singular term are given in the appendix \ref{ap:bessel}.

Velocities at the droplet interface are unknown and the provided interface conditions are the jumps in surface stresses, which we discretize from eq.(\ref{eq:laplace}).
\begin{equation}
	[\![ \mathbf{f}_i ]\!]= \frac{\gamma_{i+1/2} \frac{\bv{x}_{i+1} - \bv{x}_i}{|\bv{x}_{i+1}- \bv{x}_{i}|} - \gamma_{i-1/2} \frac{ \bv{x}_{i}-\bv x_{i-1} }{|\bv{x}_{i}- \bv{x}_{i-1}|}}{ \frac{1}{2}|\bv{x}_{i+1}- \bv{x}_{i}|+\frac{1}{2}|\bv{x}_{i}- \bv{x}_{i-1}|} - \gamma_i \left( 1-\frac{\pi}{4} \right) \frac{\frac{\bv{x}_{i+1} - \bv{x}_i}{|\bv{x}_{i+1}- \bv{x}_{i}|} - \frac{ \bv{x}_{i}-\bv x_{i-1} }{|\bv{x}_{i}- \bv{x}_{i-1}|}}{ \frac{1}{2}|\bv{x}_{i+1}- \bv{x}_{i}|+\frac{1}{2}|\bv{x}_{i}- \bv{x}_{i-1}|} +  \frac{2\gamma_i}{h}\mathbf{n},
	\label{eq:explicitstress}
\end{equation}
where $\gamma$ is defined on vertices and midpoints.

\subsubsection*{Validation: Marangoni flow of Boos and Thess}
\label{ch:marangoni}
We demonstrate the convergence of the method by comparison to an analytical solution.
A droplet placed in a Hele-Shaw cell and subject to a surface tension gradient was studied by Boos and Thess~\cite{BoosThess}. The gradient in surface tension will lead to a motion that is tangential to the interface, which is named Marangoni effect. In their theoretical study they assumed a cylindrical droplet that is exposed to a linearly changing surface tension gradient. Their result is of particular interest because it provides a test case with a shear stress boundary condition, something impossible when using the Darcy equation. The length scale $L$ is equal to the droplet radius, hence we apply boundary conditions for a droplet of radius $r=1$ and surface tension gradient in the y-direction $\frac{d\gamma}{d y} = 1$ are radial velocity $u_r = 0$ and tangential stress discontinuity $[\![ f \cdot \bv{t} ]\!] = x$. 

The $2D$~Brinkman solution by Boos and Thess is given as:
\be
	u_\theta = \frac{I_2(k) \, K_0(k) \sin{\theta}}{\lambda K_0(k) (k I_1(k) - 2 I_2(k) ) + I_2(k) (k K_1(k) + 2 K_0(k)) }.
\ee
The amplitude of the tangential velocity along the interface decreases with $1/k$ that is proportional to the channel height $h$. A high inner viscosity $\lambda$, further reduces the tangential velocity.

\begin{figure}[htb]
	\begin{center}
	\includegraphics{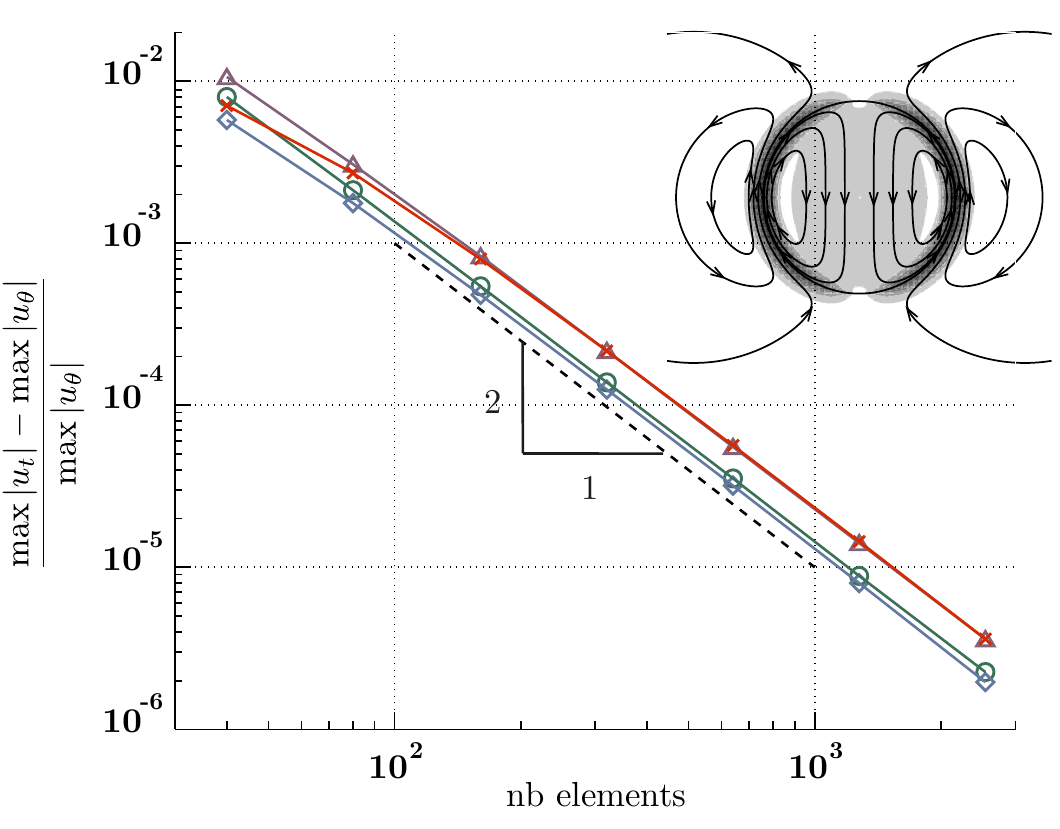}       
	\caption{Convergence for a cylindrical droplet exposed to a surface tension gradient. The error is given as $(\max|u_t| - \max|u_\theta|)/\max|u_\theta|$ and plotted against the number of elements on the circumference. $\circ$ L/H = 3, $\lambda$ = 1/3, $\triangle$ L/H = 6, $\lambda = 1/3$, $\diamond$ L/H= 3, $\lambda=5$, $\times$ L/H=6, $\lambda=5$. The inset visualizes the flow field for $L/H=3$ and $\lambda=5$.}
\label{fig:maraconv}
\end{center}
\end{figure}

Comparing theoretical velocity $u_\theta$ against numerically obtained velocity $u_t$ shows a decrease of the maximum error with second order, shown in figure \ref{fig:maraconv}. Two aspect ratios $L/H=3$ and $6$ and two viscosity ratios $\lambda = 1/3$ and $5$ were used. 

\subsection{Temporal discretization and stabilization of the fluid-fluid interface}
Two schemes are implemented to integrate the solution in time, a one step explicit Euler scheme and a two-step Runge-Kutta scheme (Heun's method).

Using the explicit Euler scheme the droplet is advanced in discrete time steps:
\begin{equation}
	d\mathbf{x} = \mathbf{u} \, dt	\quad \Rightarrow \quad \mathbf{x}^{n+1} = \int \limits_{t=n}^{t=n+1} \mathbf{u(x)} \, dt + \mathbf{x}^n \quad \Rightarrow \quad \mathbf{x}^{n+1} \approx \mathbf{u}^n \, \Delta t + \mathbf{x}^n,
	\label{eq:eulerex}
\end{equation}
Where $\mathbf{u}^n$ is the velocity field obtained by solving the boundary element problem eq.(\ref{eq:bem_discrete}) using the nodes at $\mathbf{x}^n$.

Alternatively we use a two-step Runge-Kutta scheme:
\begin{eqnarray}
	\mathbf{x}^\star = \mathbf{x}^n + \mathbf{u}^n \Delta t,
\label{eq:advance_tmp}\\
	\mathbf{x}^{n+1} = \mathbf{x}^n + \frac{\Delta t}{2}(\mathbf{u}^n+ \mathbf{u}^\star),
\end{eqnarray}
With $\mathbf{u}^\star$ being the intermediate velocity field obtained using the nodes at $\mathbf{x}^\star$.

During evolution, when the element size on the interface is twice as big or twice as small as the initial size, the points are redistributed equidistantly with the help of a $3$rd order polynomial fit.  

The time step is limited as a consequence of the mobile interface and non-linearities associated with surface tension and coupling domains of possibly different viscosities. 
In the limit of low capillary numbers when the surface tension dominates over viscous dissipation the problem evolves very slowly whereas the time step for stable integration is limited by the element size.
For instance the circular interface with constant curvature of a droplet at rest, which is physically stable, develops nevertheless divergent oscillations when solving the evolution problem for a time step larger than some factor multiplied by the discrete element length $\Delta s$. 
For a viscosity ratio $\lambda=1/2$, aspect ratio $L/H=4$ and a droplet of radius $r=1$ the maximal time step is about $\Delta t  \lesssim 8 \Delta s$.
This stiff constraint has been observed by others, e.g. Dai \textit{et al.}~\cite{shelley93} when analyzing finger formation in a Hele-Shaw cell at low capillary numbers.

\begin{figure}[htb]
\begin{center}
	\includegraphics{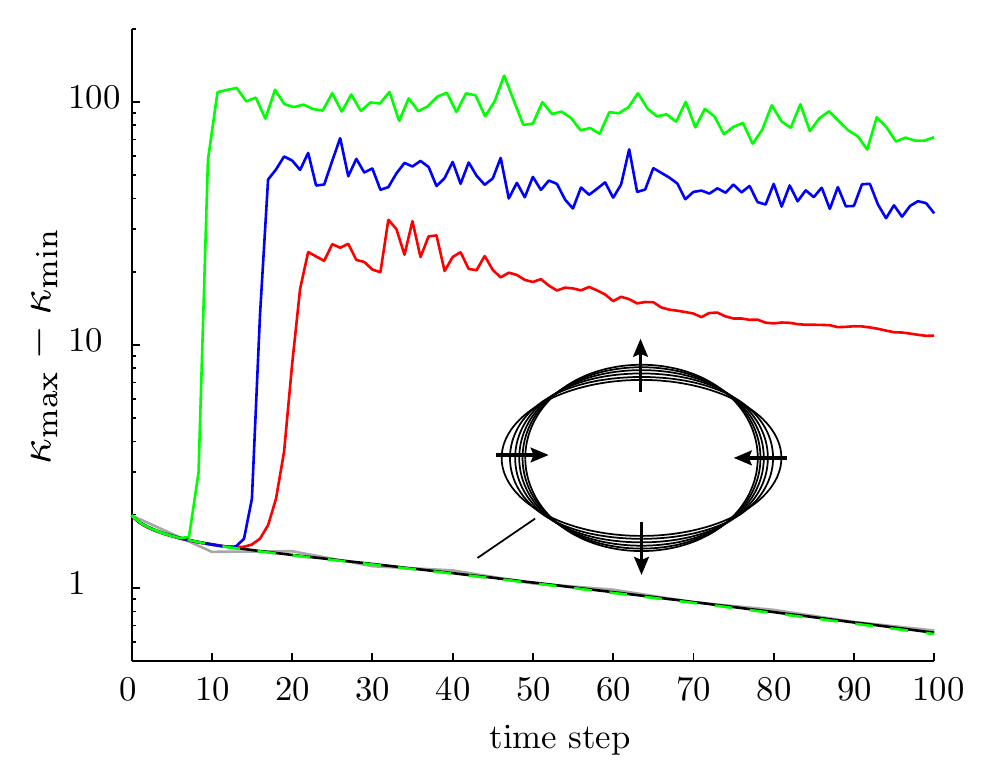}
\caption{Relaxation of a droplet with eccentricity $e=0.83$, area $\pi$, aspect ratio $R/H=4$ and viscosity ratio $\lambda=1/2$ . The black (resp. gray) line represents the stabilized scheme with $N= 400$ elements and $\Delta t=1$ (resp. $\Delta t=10$). The non-stabilized scheme is represented by the green line $N=400, \Delta t= 0.125$, blue $N=200, \Delta t = 0.25$, red $N =100, \Delta t= 0.5$ and green dashed line $N = 400, \Delta t=0.0625$.}
\label{fig:ellgraph}
\end{center}
\end{figure}

The numerical instability is illustrated here on a dynamic problem, where a droplet whose initial shape is an ellipse of eccentricity $e=0.83$ and area $\pi$ relaxes to a circular interface. 
A discretization of 100, 200 and 400 elements is tried with time steps on the threshold of instability of $\Delta t$ = 8 $\Delta s$, so $\Delta t = 0.5, 0.25$ and $0.125$. 
In figure \ref{fig:ellgraph} the difference between maximal and minimal in-plane curvature is plotted, which starts near two and goes to zero for the perfectly cylindrical droplet. 
In plotting the curvature, oscillations will appear much more pronounced. 
The time steps on the threshold lead to oscillations due to deformations on the evolving interface. As the time step is divided by two the scheme becomes stable, as shown by the green dashed line with $N=400$ and $\Delta t = 0.0625$.

Hou \textit{et al.} \cite{shelley94} developed a linearization technique to remove the stiffness from surface tension but their approach requires a transformation of the equations in terms of tangential and angular coordinates (s,$\theta$).
We propose a different approach where the curvature is also linearized, however the variables remain in cartesian coordinates.

\subsubsection*{Interface stabilization technique}
A semi-implicit scheme is used to stabilize these numerical capillary waves and therefore allow for larger time-steps. The scheme used here in two dimensions extends in a straight forward manner to three dimensions.

For a discrete interface the in-plane interface stress is derived from eq.(\ref{eq:explicitstress}) at the vertex $i$ at time step $n$ as:
\begin{equation}
	\frac{d \gamma \bv t}{d s} \Big|_i^{t=n} = \frac{\gamma_{i+1/2} \frac{\bv{x}_{i+1} - \bv{x}_i}{|\bv{x}_{i+1}- \bv{x}_{i}|} - \gamma_{i-1/2} \frac{ \bv{x}_{i}-\bv x_{i-1} }{|\bv{x}_{i}- \bv{x}_{i-1}|}}{ \frac{1}{2}|\bv{x}_{i+1}- \bv{x}_i|+\frac{1}{2}|\bv{x}_i- \bv{x}_{i-1}|} \bigg|^{t=n}.
	\label{eq:implicitcurvature}
\end{equation}

Stabilization is achieved when the interface is "made aware" of its displacement when providing a feedback loop. 
The position of the points $\bv x^\star$ a discrete time step $\Delta t$ after time $n$ is approximated by eq.(\ref{eq:advance_tmp}) to be $\bv x^\star = \bv{u}^n+\Delta t \; \bv{u}(\bv{x}^{n})$. Presuming the change in distance between two points to be negligible ($|\bv{x}_{i+1}^{\star}-\bv{x}_{i}^{\star}| \approx |\bv{x}_{i+1}^n-\bv{x}_{i}^n|$), the linearized expression for the interface stress after a time increment $\Delta t$ is then given by:

\be
	\frac{ d \gamma \bv t}{d s}\Big|_i^{\star}= 
	\frac{ d \gamma \bv t}{d s}\Big|_i^n
	+
	\frac{\gamma_{i+1/2} \frac{\bv{u}_{i+1} - \bv{u}_i}{|\bv{x}_{i+1}- \bv{x}_{i}|} - \gamma_{i-1/2} \frac{ \bv{u}_{i}-\bv u_{i-1} }{|\bv{x}_{i}- \bv{x}_{i-1}|}}{ \frac{1}{2}|\bv{x}_{i+1}- \bv{x}_{i}|+\frac{1}{2}|\bv{x}_{i}- \bv{x}_{i-1}|} \Delta t \bigg|^{t=n}.
\ee

A similar stabilization is applied to the second term in eq.(\ref{eq:explicitstress}).
Using the stabilization scheme for the relaxing droplet allows for much larger time step.
Computing the droplet relaxation with 400 nodes and a time step of $\Delta t=1$ and $10$ agrees with the non-stabilized solution that was computed at a time step that is 16 and 160 times smaller, respectively. 
For the stabilized scheme no oscillations are observed and stabilization is achieved independent of the discretization. 

\subsubsection*{Convergence study: Deformable droplet in flow focusing}
\label{ch:convcheck}
\begin{figure}[htb]
	\begin{center}
	\includegraphics[scale=1]{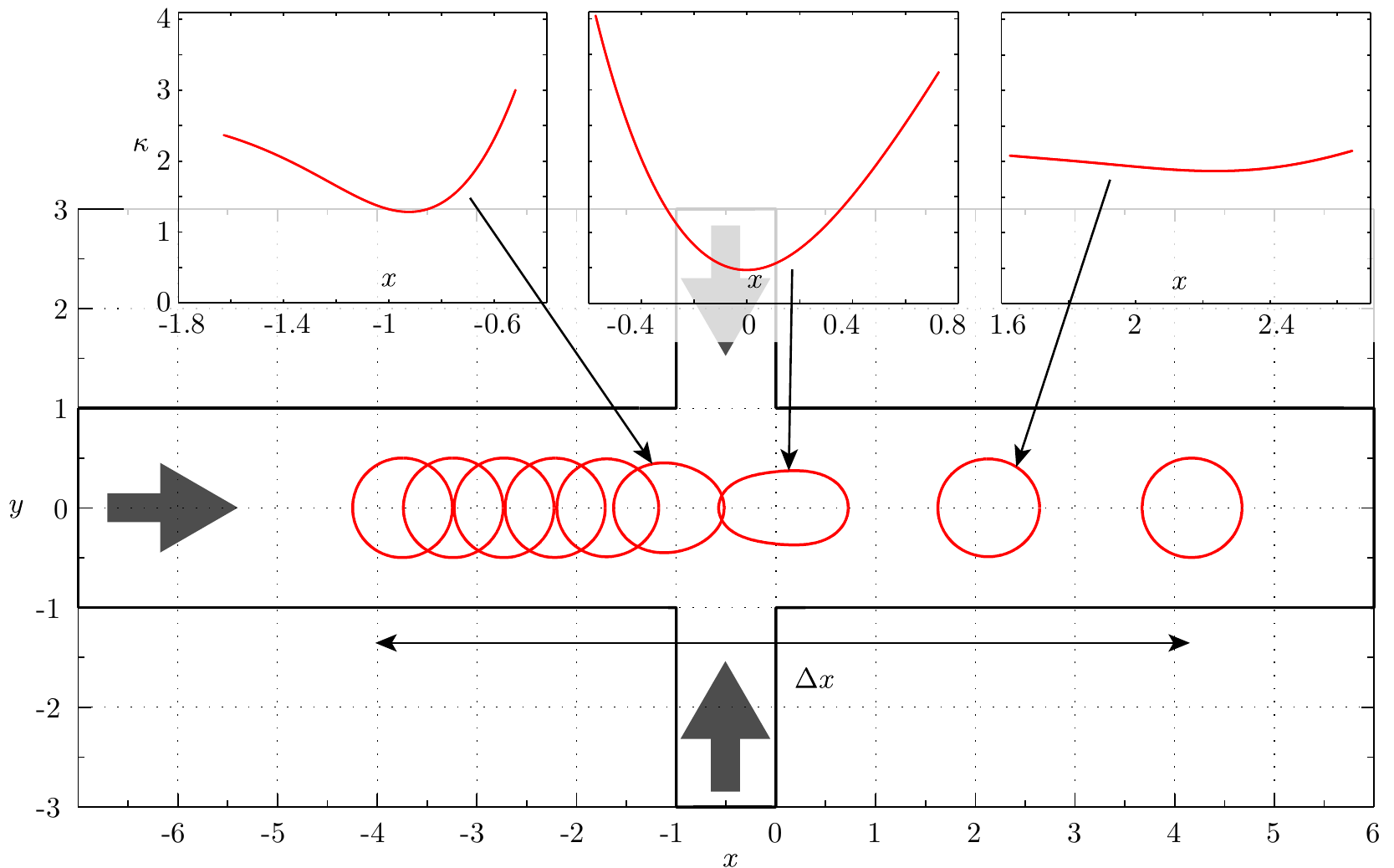}       
	\caption{The figure shows the computational domain with the droplet interface at several time steps. The three inset figures show the in-plane curvature at three selected time steps around the most deformed. The interface is free of any spurious oscillations.}
\label{fig:flowfocus}
\end{center}
\end{figure}

In lack of an analytical solution for the convergence study of deformable droplets we resort to comparison with a numerical result of increased spatial and temporal resolution. 
For this study a cross flow junction is simulated, where a fluid stream is focused by flow from two side channels. 
A droplet of diameter $r=\frac{1}{2}$ submitted to that flow is deformed and accelerated, figure \ref{fig:flowfocus} shows the numerical set-up with time lapsed droplet position. The left inflow condition is chosen to be $Ca=0.005$ and the inflow velocity from the side channels is $Ca= 0.015$.

Droplet viscosity is half the viscosity of the surrounding fluid, $\lambda=\frac{1}{2}$ and droplet aspect ratio $L/H = 3$. Variables that are observed are the displacement $x_c$ of the center of mass of the droplet and its perimeter. 

Displacement error is measured by the root-mean-square (rms) of the difference in position at $n=68$ time steps normalized by the total distance. The perimeter error is the rms of the difference of the droplet perimeter, sampled at $68$ time steps and normalized by the perimeter for an undeformed droplet $\pi$.

Spatial discretization varies between $\Delta x = \frac{1}{100}, \cdots, \frac{1}{10}$ as element size for the walls and $\Delta x = \frac{1}{300}, \cdots ,\frac{1}{30}$ for the droplet  to account for the importance of the mobile interface. Temporal discretization ranges between $\Delta t = \frac{10}{16} , \cdots , 10$. The reference solution uses $\Delta t = \frac{1}{10}$ and $\Delta x = \frac{1}{200}$ and $\Delta x_{\textrm{drop}} = \frac{1}{600}$, respectively.

\begin{figure}[htb]
	\begin{center}
	\includegraphics[scale=1]{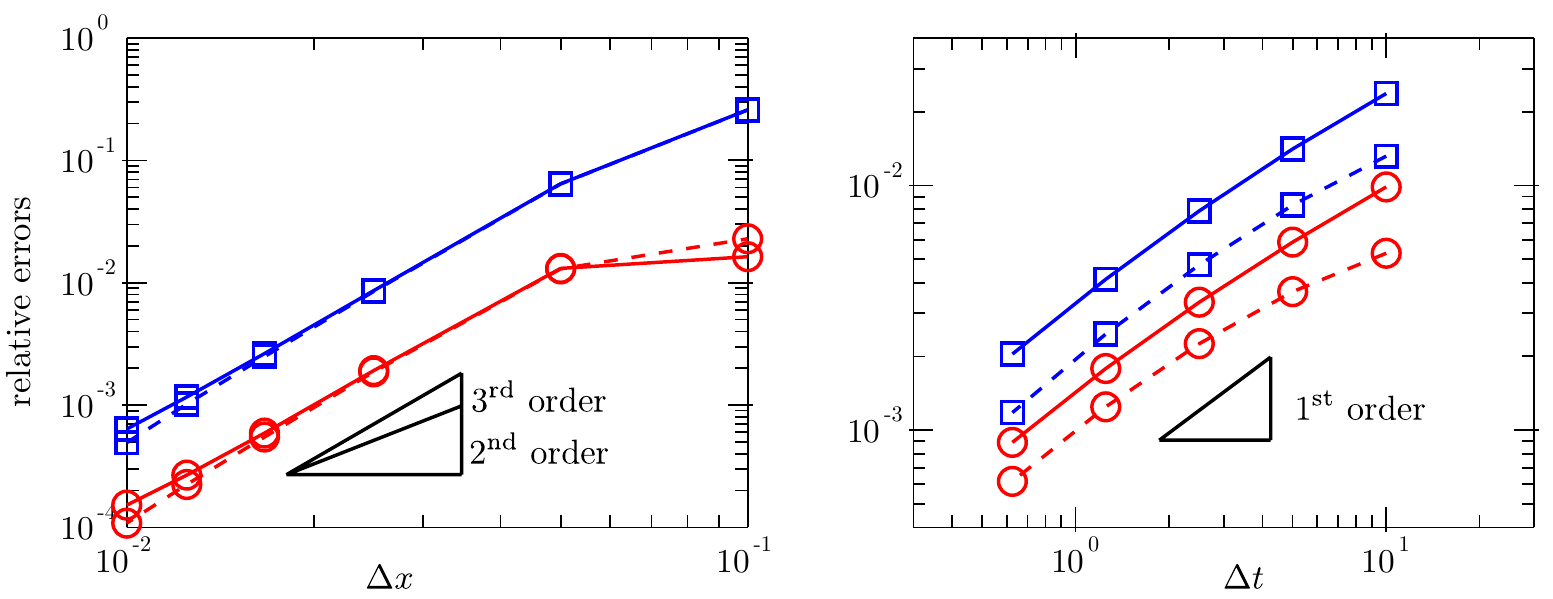}       
	\caption{Left figure a) convergence for varying element size $\Delta x$, right figure b) convergence for varying time step $\Delta t$. The blue line represents the displacement error $\epsilon_x$ and the red line the perimeter error $\epsilon_p$. Full line for 1-step integration and dashed line for 2nd order Runge-Kutta scheme.}
\label{fig:dynamic_conv}
\end{center}
\end{figure}

Figure \ref{fig:dynamic_conv} a) shows the errors for a fixed temporal resolution of  $\Delta t = \frac{1}{10}$ and confirms second order convergence as expected from the previous study in section \ref{ch:marangoni}. 
In fact the error decreases almost with 3rd order, which maybe due to the symmetries in the configuration.
Figure \ref{fig:dynamic_conv} b) shows for a fixed spatial resolution of $\Delta x = \frac{1}{200}$ the temporal convergence. 
The convergence of the stabilized Euler scheme is of order $1$ as expected. However the second order Runge-Kutta scheme (Heun's method) shows also first order convergence although with smaller error, which is likely due to the fact that an intermediate time step was performed. The semi-implicit interface stabilization scheme incorporates a first order Euler scheme and thus spoils any higher order scheme.

Without interface stabilization the spatial convergence, done with $\Delta = \frac{1}{10}$ would fail for $\Delta x< 0.04$ on the droplet, which follows from the empirical formula above $\Delta t  \lesssim 8 \Delta s$. Likewise the temporal convergence with fixed spatial discretization of $\Delta x=\frac{1}{200}$ would require a time step as low as $0.013$ that is eight times lower than the time step in the refined solution.

Due to incompressible flow the area of the droplet, initially $A_0=\pi/4$ is theoretically conserved. In the worst case, the scheme with the lowest resolution finished the simulation loosing $3.4\%$ of the area, whereas the scheme with the highest resolution lost $8 \cdot 10^{-4}\%$.

\subsection{Block pre-elimination and parallel scaling}
\label{sec:blockelim}
The discretized problem results in a dense matrix $\bv{A}$ with right-hand-side $\bv{b}$ and vector of unknowns $\bv{u}$, containing velocities and stresses. The degrees-of-freedom (DOF) of the problem are $2(M+N)$. $2M$ DOF associated to the static interface, the outer walls, and $2N$ DOF associated to the dynamic interface, the droplet. If the droplet is surrounded by walls one generally finds that the DOF, $2M$, of the walls are larger than the DOF, $2N$, of the droplet. The problem without droplet is without evolution and independent of time. One therefore splits the matrix into four blocks:
\begin{equation}
	\mathbf{A} \mathbf{u} = \mathbf{b}  \Rightarrow 
 \begin{pmatrix}
  \mathbf{W}  & \mathbf{R}  \\
  \mathbf{P} & \mathbf{D} 
 \end{pmatrix}
  \begin{pmatrix}
  \mathbf{u}_w  \\
  \mathbf{u}_d
 \end{pmatrix}
 = 
 \begin{pmatrix}
 \mathbf{b}_w  \\
 \mathbf{b}_d
 \end{pmatrix}
 \label{iproblem}
\end{equation}
Here $\mathbf{W}$ is the influence of the static walls on themselves, $\mathbf{R}$ like a resistance, is the influence of the droplet on the outer walls, $\mathbf{P}$ like propulsion, is the influence of the outer walls on the droplet and $\mathbf{D}$ the influence of the droplet on itself. 

Matrix sizes are $\mathbf{W}$ is $4 M^2$, $\mathbf{R}$ and $\mathbf{P}$ are $4 MN$ and $\mathbf{D}$ is $4 N^2$. The matrix $\mathbf{W}$ is always the same because it's boundary conditions and its element distribution does not change. We invert the matrix and save the inverse $\mathbf{W}^{-1}$.

Applying the inverse to the upper part of the eq. \ref{iproblem}:
\[
	\mathbf{W} \mathbf{u}_w + \mathbf{R} \mathbf{u}_d = \mathbf{b}_w \quad \Rightarrow \quad \mathbf{u}_w = -\mathbf{W}^{-1} \mathbf{R} \mathbf{u}_d -\mathbf{W}^{-1} \mathbf{b}_w.
\]
Then replacing $\mathbf{u}_w$ in the lower part of eq. \ref{iproblem}:
\[
	\mathbf{P u}_w + \mathbf{D u}_d = \mathbf{b}_d \quad \Rightarrow \quad -\mathbf{P W}^{-1} \mathbf{R u}_d - \mathbf{P W}^{-1} \mathbf{b}_w + \mathbf{D u}_d = \mathbf{b}_d .
\] 
And finally:
\be
	(\mathbf{D-P W}^{-1} \mathbf{R) u}_d = \mathbf{P W}^{-1} \mathbf{b}_w + \mathbf{b}_d.
	\label{eq:schur}
\ee
With $\mathbf{D-P W}^{-1} \mathbf{R}$ called the Schur complement. These matrix-matrix and matrix-vector multiplications reduce the problem from a dense $4(M+N)^2$ matrix to a dense $4N^2$ matrix. In the case of equal viscosities $\lambda=1$, $\mathbf{R} = \mathbf{0}$ and $\mathbf{D}$ is the identity matrix, so the velocity at the droplet interface is given explicitly by: $\mathbf{u}_d = \mathbf{P W}^{-1} \mathbf{b}_w + \mathbf{b}_d$.

The described boundary element code has been implemented in the C++ programming language. Independently of the block pre-elimination we have used shared memory parallelization.
A direct matrix solver for general matrices implemented in {LAPACK}~\cite{lapack} solves the matrix. Dense matrices in contrast to sparse matrices have stronger limitations for parallelism on distributed computers.
A rather efficient way is to use OpenMP\cite{openmp}, which allows shared memory parallelism on a multiprocessor and multicore environment. LAPACK routines automatically use these features, whereas the loop for matrix filling has been parallelized by OpenMP pragma.

At each iteration in time, matrix filling is done by nested loops. An outer loop over all boundary elements and a nested loop that integrates over all Greens functions at the collocation points, where the numerical integration is a further nested loop. This inner most loop over only six points is too small for efficient parallelization. Hence the loop over all Greens functions is parallelized.

\paragraph*{Scaling for parallelism and pre-condensation}
The speed-up is demonstrated solving for a droplet advected in a rectangular channel using pre-condensation and multiple cores. The boundaries were discretized with 1440 DOF and its multiple by five and ten (7200 and 14400 DOF). The ratio between DOF on the droplet and on the fixed geometry, $M/N$ was changed between $3,5,9,14$ and $19$.  

\begin{figure}[htb]
\begin{center}
	\includegraphics[scale=1]{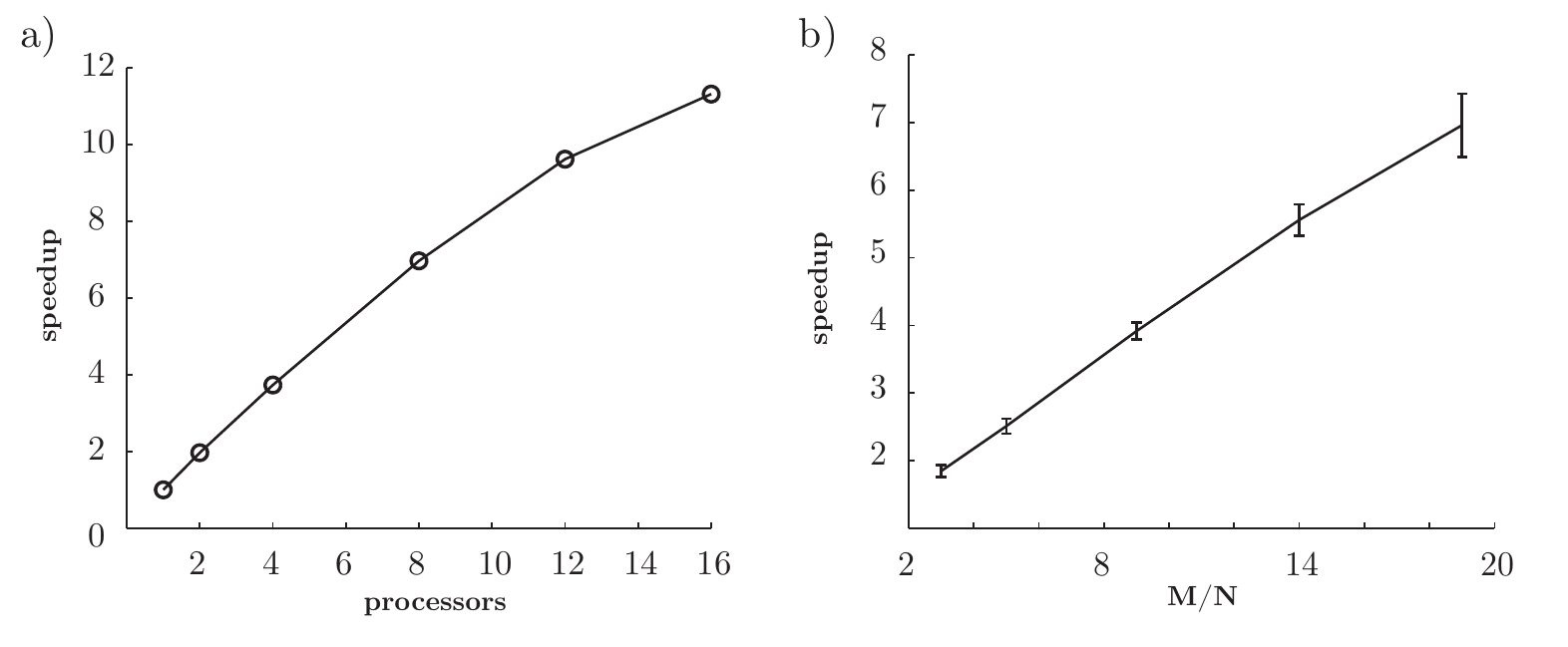}
\caption{Speed-up of the simulation by: a) OMP implementation, where the speed-up increases almost linearly to the number of cores, mean of $7200$ and $14400$ DOF shown. b) Pre-condensation, where the speed-up increases approximately linearly with the ratio of droplet to wall degrees-of-freedom.}
\label{fig:performance}
\end{center}
\end{figure}

The performance tests were performed on a Dell Server with 16 cores at 1.8 GHz. The computation time for a single iteration of 1440 DOF on a single core was $0.57 s$ with 7200 DOF $25.91 s$ and with 14400 DOF $161.62 s$. Practically all the time is spent on matrix filling and matrix solving. There is an almost perfect scaling of computation time $T \propto \textrm{DOF}^2$ for matrix filling and $T \propto \textrm{DOF}^3$ for matrix solving.

In the presence of outer walls the fixed boundary problem can be pre-condensed and therefore reduces the time to fill and solve the matrix. In assessing the speed-up we ignore the time associated to pre-condensation, which includes one matrix inversion. The matrix inversion is more costly than direct solving of a matrix but it is done only once and becomes negligible to the overall time. The larger the fixed part of the matrix the higher the speed-up.

Increasing the number of cores shows a good scaling up to 16 cores for the problems with $7200$ and $14400$ unknowns, which achieve a speed-up of about $11$ compared to the single core case. The parallelism work about equally well for matrix filling and solving.
The problem with $1440$ unknowns shows more waiting time and achieves only a speed-up of 7. We shall exclude the results so the trends are uniform but apply only to the large scale problems, which are in fact the ones that have the highest need for acceleration. Figure \ref{fig:performance} a) shows the speed-up time on a single core divided by time on multiple cores.

Comparing problems of varying ratio $M/N$ in figure \ref{fig:performance} b) shows the mean speed-up out of four configurations, from $7200$ DOF on single core and 16 cores and $14400$ DOF on a single core and 16 cores. The error bars are computed from configurations different DOF with and without parallelism and indicate that the speed-up is quite uniform also when using OpenMP. The problem with $1440$ DOF was stagnant at a speed-up by factor 3.

\section{Applications compared to theory and experiments}
\label{results}
The physical significance of the presented method is demonstrated on three cases: 
Comparison to a linear stability analysis of the Saffman-Taylor instability and two experiments on droplet deformation.

\subsection{The Saffman-Taylor instability}
Finger formation as a consequence of the Saffman-Taylor instability is an archetypical problem of dynamically developing interfaces. 
It has been studied numerically using Darcy's law for instance by Dai \textit{et al.}~\cite{shelley93} who looked at the amplification of small perturbations on the unstable interface.

The instability and resulting finger formation is due to a viscosity change over an interface, where a less viscous fluid drives a more viscous fluid. 
In opposition, surface tension damps surface perturbations that would cause highly curved regions. For a radial configuration the growth rate using the Brinkman equation was derived recently~\cite{nagela} and showed better agreement than Darcy's law for relatively large capillary numbers.

Introducing a sinusoidal perturbation of the interface, the growth of the perturbation is given as $a(t) = a_0 \exp(\omega t)$. The initial aspect ratio, radius over height, was set to $L/H=80$, the viscosity ratio to $\lambda= \frac{1}{10}$ and the capillary number was varied between $Ca =1, 0.03$ and $0.01$ as to obtain unstable, neutral and stable modes. The instability is numerically simulated with a droplet interface of $500$ elements and a time step of $\Delta t=10^{-3}\cdot Ca^{-1}$.

\begin{figure}[htb]
\begin{center}
	\includegraphics[scale=1]{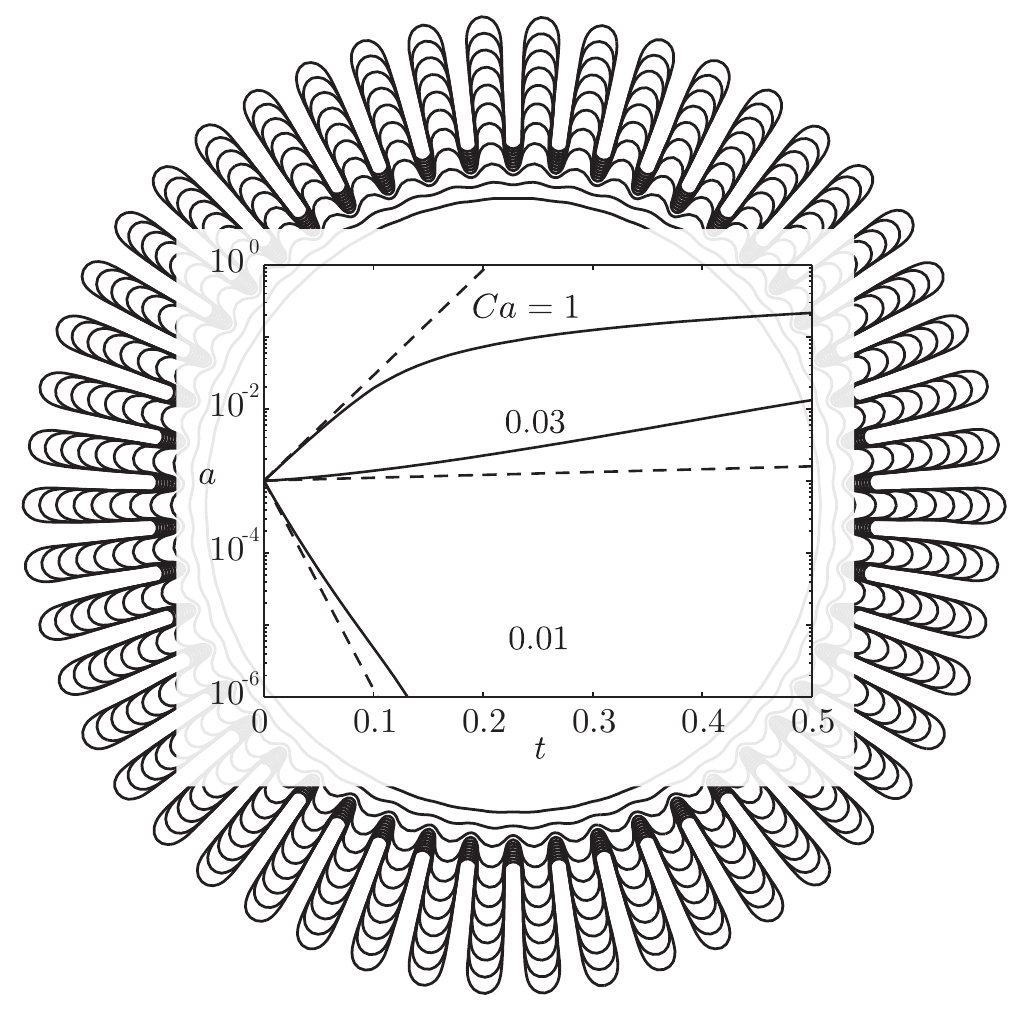}   
	\caption{Plotting the growth of a sinusoidal perturbation of wave number $50$ while injection of a liquid that is $10$ times less viscous than the surrounding liquid. The evolution of the interface for $Ca=1$ is depicted at several instances in time. The graph in the center compares prediction by linear stability analysis - - - against numerical simulation ---.}
	\label{fig:saffman}
\end{center}
\end{figure}

Plotting perturbation amplitude $a(t)$ against time for analytical and numerical results $a^n = \frac{1}{2}(R^n_{max} - R^n_{min})$ in figure \ref{fig:saffman} shows perfect agreement in the beginning of the simulation, then the simulation is slightly parting from the prediction as the radius evolves. The background illustration in the figure shows the interface evolution for every $50^\textrm{th}$ time step. Deviation for large $t$ at $Ca=1$ is due to non-linear saturation and at $Ca=0.03$ because the aspect ratio increases with the radius and therefore the solution shifts into the unstable regime.

When starting from stochastic initial conditions these simulations have allowed to  observe tip splitting and opens perspectives to study the selection of number of fingers.
When adding Marangoni stresses it is also possible to analyze the evolution of fingering in a medium of variable surface tension, like in a Hele-Shaw with thermal gradients.

\subsection{Comparison with experiments: Droplet stretching}
Two different cases of droplet stretching in microchannels are presented; first at large and then at low aspect ratio. 

\paragraph*{Stretching in a hyperbolic flow}
Recently Ulloa \textit{et al.}~\cite{cordero14} experimentally studied droplet deformation in diverging flow at low Reynolds numbers and aspect ratio of droplet radius to channel height $R/H>1$.  The viscosity ratio of inner fluid to outer fluid is $\lambda=0.008$ in all their experiments as well as in our simulations. Their set-up is a microfluidic cross-junction with fluid injected from two opposite sides, where the fluid leaves the junction in two channels, oriented $90^\circ$ to the inlet channels.

From left inlet channel droplets enter the junction and get advected to the center of the junction, a stagnation point, and stretched by the flow. 
This stagnation point is a saddle point and due to imperfect alignment and the unstable nature of the equilibrium position the droplet gets streamed away after some time.
Before getting streamed away the droplets have deformed in the flow field to an equilibrium shape. Ulloa \textit{et al.} \cite{cordero14} characterized the deformation $D$ by the difference in major and minor axis length $a$ and $b$, normalized by their sum, $D = (a-b)/(a+b)$. They investigated the influence of channel geometry, shear rate and droplet radius on the deformation.

\begin{figure}[htb!]
\begin{center}
	\includegraphics[width=\textwidth]{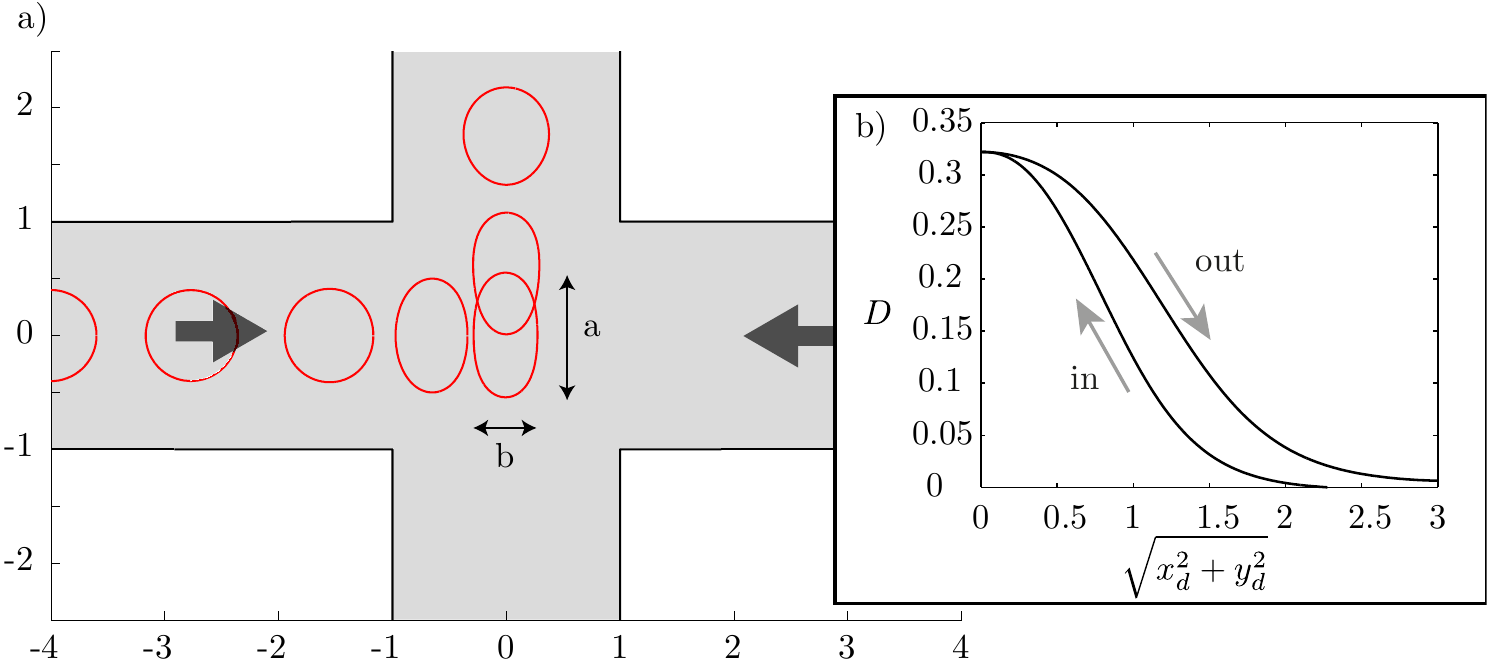}   
	\caption{Multiple snapshots of a droplet that enters a junction from the left. Gray arrows indicate the incoming flow. Capillary number $Ca=0.048$, channel width $W/H=7.5$ and droplet radius $2R/W=0.4$. The inset b) shows droplet deformation, plotted against the distance between droplet center and the center of the junction. }
	\label{fig:junction}
\end{center}
\end{figure}

\begin{figure}[htb!]
\begin{center}
	\includegraphics[width=\textwidth]{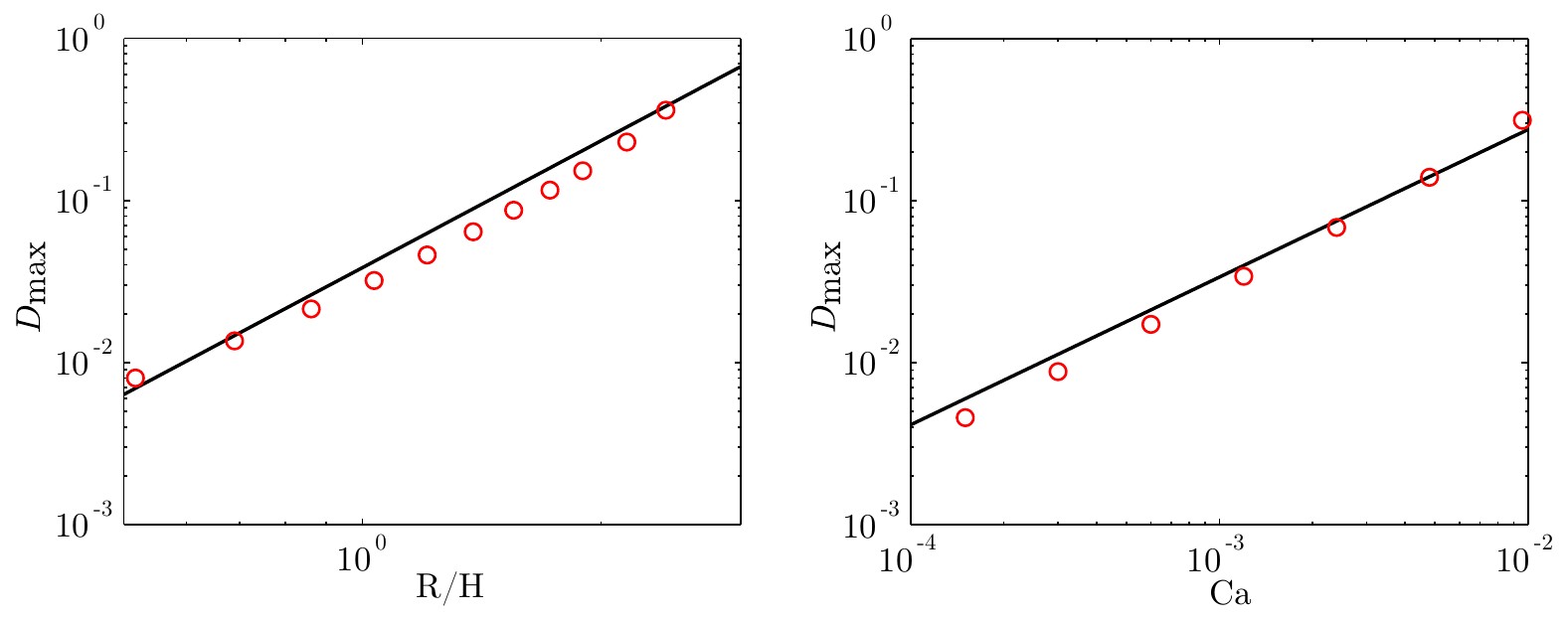}   
	\caption{Maximum deformation $D_\textrm{max}$ against a) droplet aspect ratio $R/H$ at capillary number $Ca =0.0126$ and channel aspect ratio $W/H = 6.9$ and b) against capillary number $Ca$ for droplet aspect ratio $R/H=1.5$ and channel aspect ratio $W/H=7.5$. The black line is the trend extracted from the experimental data and the circles correspond to simulations.}
	\label{fig:fig4compare}
\end{center}
\end{figure}

Choosing that channel half width as a length scale $L=W/2$, a junction of width $2$, extending from $-6$ to $6$ in the $x$-direction and $-4$ to $4$ in the $y$-direction is simulated. For illustrative purposes figure \ref{fig:junction} shows a simulation with one droplet at several instances. One sees the droplet coming from the left, stretching in the center and leaving the junction on the top. A maximum deformation of $D=0.323$ is reached at the center as shown in the inset.

In figure \ref{fig:fig4compare} we compare the scaling of maximum deformation $D_{\textrm{max}}$ with droplet radius normalized by the channel height $R/H$ in a channel of width $W/H = 6.9$, at constant capillary number $Ca=0.0252$. Furthermore we compare to the shear rate $G$ that has been non-dimensionalized by the viscosity, surface tension and channel width $Ca= \frac{\mu_{c} G W}{\gamma_\textrm{ref}}$ in a Channel of $W/H=7.5$. Numerical simulation shows a dependence in $R/H^{2.57}$ and in $Ca^{1.01}$. Experimentally Ulloa \textit{et al.} found $D \propto (R/H)^{2.59} Ca^{0.91}$.

\paragraph*{Stretching in extensional flow}
In a recent work Brosseau \textit{et al.}~\cite{barets} investigated the influence and adsorption of surfactants. They follow the idea of a microfluidic tensiometer, which was brought up by Cabral and Hudson~\cite{cabral}, who proposed to measure the surface tension by observation of droplet deformation in an extensional flow created in a micro-channel. 

Influence of surfactants is difficult to determine experimentally and microfluidics might help to investigate characteristics that are inaccessible by established means. When surfactants are absorbed on the fluid interface they change the surface tension and introduce surface tension gradients. Modeling surfactants in numerical simulations is a formidable task and the flow solver we developed is able to simulate deformable interfaces and surface tension gradients, while being computationally inexpensive as to run many simulations in order to fit surfactant models to experiments by reverse engineering.
In the adsorption limited regime no convection-diffusion equation needs to be solved as the surfactants are spread homogeneously over the domain. In contrary in the diffusion limited regime an advection diffusion equation needs to be solved in addition to the Brinkman equation.

\begin{figure}[htb]
\begin{center}
	\includegraphics[width=\textwidth]{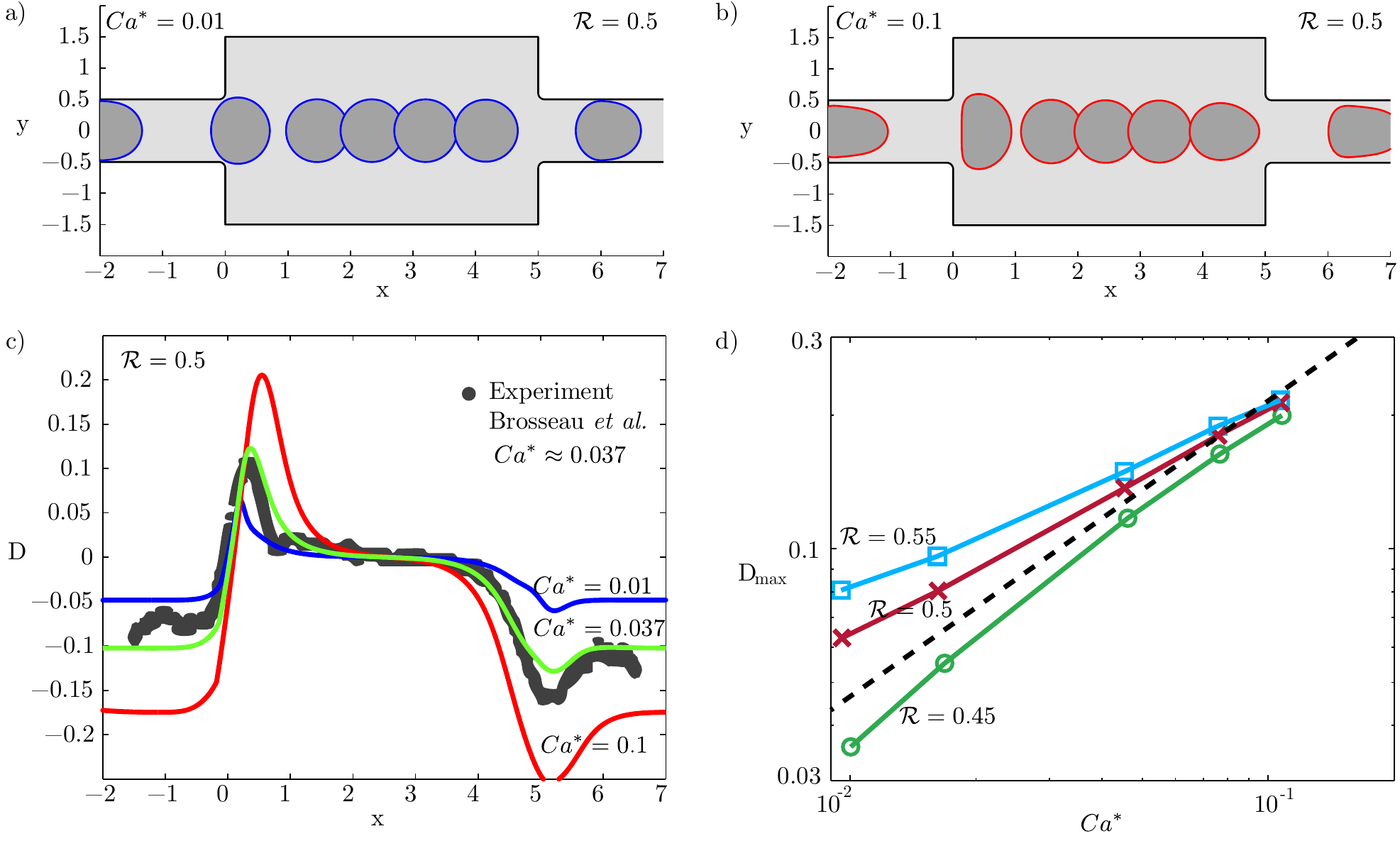}   
	\caption{a) Droplet stretching at capillary number $Ca^*=0.01$ and b) at $Ca^*=0.1$; Droplet contours at several moments in time are plotted. Graph c) shows the evolution of the deformation at different positions of the droplet. Grey dots show the experimental data from Brosseau \textit{et al.} for $\mathcal{R}=0.5$ and $Ca^*=0.037$. In figure d) the maximum deformation is plotted for different aspect ratios $\mathcal{R} =\bigcirc \, 0.45,\; \times \, 0.5$ and $\square \, 0.55$. Experimental results by Brosseau \textit{et al.} for $\mathcal{R} \approx 0.5$ are drawn by a dashed line.}
	\label{fig:brosse}
\end{center}
\end{figure}

Here we compare the first series of experiments of Brosseau \textit{et al.} \cite{barets} without surfactants to our simulations. The channel they used has a constant height of $100\mu m$ and has wide section, $300 \mu m$ wide and $500 \mu m$ long. This wide section has an entry and an exit channel $100\mu m$ wide. 
For non-dimensionalization the length scale is the entry channel width $L=100\mu m$, viscosity ratio is $\lambda = 0.8$ and channel aspect ratio $L/H=1$, as in the experiments. Simulation of square channels are out of the nominal range of validity of the presented Brinkman model. Nevertheless the value of this comparison is to estimate the limitations of our method when approaching low aspect ratios.

The wide channel section is $5$ units long and $3$ units wide and joined by $1$ unit wide entry and exit channels. The edges where these channels join have been rounded with a radius of $1/10$ that can also be observed in the experimental setup. Three different droplet aspect ratios were used $\mathcal{R}=R/L=0.45, 0.5$ and $0.55$. 
 
The experimental study used the droplet velocity in the entry channel to set the capillary number, which we shall denote here $Ca^*=\frac{\mu U_D}{\gamma}$.
Illustration of the extensional flow together with the resulting deformation for $Ca^*=0.01$ and $Ca^*=0.1$ and radius $\mathcal{R}=0.5$ is shown in figure \ref{fig:brosse} a) and b). 
At capillary number $Ca^*=0.1$ the droplet is more stretched in the entry channel, till $x=0$ and then also expands much wider than the droplet at $Ca^*=0.01$.
In figure \ref{fig:brosse} c) one can see this behavior in the deformation curve, where the initial deformation is negative, which corresponds to elongation. The deformation $D$ is measured by the difference in major and minor half axes $a$ and $b$, $D = (a - b)(a + b)$. At $x=0$ the flow expands and shortly after the droplets reach their maximum deformation. Results are shown for the examples a) and b) as well as for a comparison with a given experimental data set, showing a good agreement.

The maximum deformation is shown in figure \ref{fig:brosse} d) against the capillary number for radii $\mathcal{R} = 0.45, 0.5$ and $0.55$. The experimentally obtained scaling law between deformation and capillary number $D_{\max} \propto {Ca^*}^{2/3}$ is only approximately retrieved for $\mathcal{R} = 0.45$ and $0.5$, with a root-mean-square error of about {40\%}.

It shall be stressed that although the channel aspect ratio $L/H=1$ is not in the regime of microchannels with large aspect ratios qualitative agreement could be achieved. Furthermore the Reynolds number was beyond the domain of validity of the Brinkman equation. The Reynolds number being estimated based on the properties of the carrier fluid (fluorinated oil: $\rho = 1648 kg/m^3, \mu = 1.24 m Pa\,s$), channel height $100\mu m$ and maximum velocity of the droplet $300 mm/s$, yields a maximum value of $Re \approx 40$. Despite the relatively large values no dependence on the Reynolds number could be established from the experimental data.

A simulation of droplets with lateral confinement, as shown here, would not have been possible with the Darcy equation because without the in-plane Laplacian no coating boundary layers build up at the lateral boundaries of the droplet, resulting in collision of the droplet interface with the lateral walls.

\section{Conclusion}
\label{conclusion}
The boundary element method that is presented here solves for two-phase flows in geometries of moderate aspect ratio. Convergence of the numerical scheme has been verified as well as functionality of an interface stabilization scheme and acceleration due to Gauss block pre-condensation.

The dynamic evolution of the interface in time has been verified with the dispersion relation for the Saffman-Taylor instability based on the Brinkman equation. The perturbation amplitude of the linearized equations has been compared to the numerically obtained amplitude and results agree very well showing that the solver preserves stable, neutral and unstable modes in a non-linear problem

The deformation of a droplet in a linear flow in a wide channel at low Reynolds number agreed well with experimental findings, where the droplet aspect ratio and capillary number were varied.
In a complementary study the validity of the $2D$~model equation was checked in the limit when channels have a square cross section, where droplets deform in a sudden expansion.
Main results as the deformation over position diagram closely resembles the data obtained in experiments, as well as the scaling of the maximum deformation over capillary number are recovered. However the $2D$~model breaks down at low capillary numbers, where the surface tension forces the droplet interface very close to the walls.

Furthermore only flows in the absence of inertia can be solved by the boundary element method without resorting to domain integration, imposing a Reynolds number $Re \ll 1$, whereas in some cases microfluidic experiments have Reynolds number larger than $Re=1$. Domain integration or dual reciprocity methods \cite{pozrikidis} can be used to include these terms, but they introduce a considerable effort compared to the presented procedure. In the experimental study at finite Reynolds number \cite{barets} to which we compared our simulations, the results seemed to be independent of the Reynolds number.
An explanation could be that the Weber number, which measures the inertial terms compared to the surface tension $W = \frac{\rho U^2 L}{\gamma}$ remains small, $We \approx 1$ at most in the experiments of Brosseau \textit{et al.}. Hence the surface tension might be expected to dominate over inertial terms and the dynamics of the interface become not too much affected.

Given the restrictions of being a $2D$~approach with domain wise constant parameters it should be stressed that the method is rather fast. Simulations run on a desktop computer typically finish in a couple of minutes or hours depending on the problem size, which in return allows doing extensive parametric studies and simulating a large number of droplets.

The acceleration is due to the reduced number of degrees of freedom when using BEM but also to acceleration techniques and the interface stabilization. 
The latter is important since the droplet evolves on a capillary time scale that is $T= L \mu_c / \gamma_{ref}$, which for typical values like $L = 100 \mu m, \mu_c = 10^{-3} Pa\,s $ and $\gamma_{ref} = 10^{-2} Pa\, m$ gives $T = 10^{-5}s$. Considering for instance an element size of $80$ elements per unit length and a admissible time step $\Delta t= 8 \Delta s$ it will take about $10^6$ iterations to advance $1s$ in physical time. Applying stabilization allows for $\Delta t > 1$ with no restriction on the spatial discretization.
Without any kind of interface smoothing scheme the interface remains oscillation free.
The relatively low computational effort comes to a price: Depth-averaging is based on a high aspect ratio and the soundness of the results becomes questionable when the droplets are not sufficiently confined. 

The variety of problems that we solved demonstrates that the tool has the capability to simulate the flow of deformable droplets in complex channel geometries encountered in Lab On A Chip applications. 
In two recent publications we applied the solver to study droplet relaxation~\cite{nagel2} and droplet trapping~\cite{nagel3} and have obtained good agreement between simulation and experiments. 
Current work includes modified boundary conditions that take into account the effect of film formation by coupling the interface boundary condition to an asymptotic solution that modifies the out-of-plane curvature according the asymptotic developments of Park and Homsy~\cite{park84}.

The fact that the physical description and numerical algorithm is reduced to its essentials makes it relatively easy to implement and efficient to run. Therefore this work could provide a further step towards numerical simulation of droplet microfluidics and therefore enable experimentalists to use simulations to design or evaluate their Lab-on-a-Chip set-up.

\section*{Acknowledgments:}
We thank the European Research Council for funding this work through a starting grant (ERC SimCoMiCs – 280117).

\appendix

\section{Numerical approximation of Bessel functions}
\label{ap:bessel}
The modified Bessel functions of second kind are developed as a series.
The modified Bessel differential equation is:
\[
	\frac{d^2 y}{dx^2}x^2 + \frac{dy}{dx} x- y(x^2+n^2) = 0,
\]
Whose solutions are the modified Bessel functions of second kind $I_n(x)$ and $K_n$, where we are only interested in $K_0$ and $K_1$. The differential equation has two singular points, one at $x=0$ and another at $x=\infty$. Therefore the functions are developed in two zones around those points and meet at $x=2$. An estimation of the maximum error around $x=2$ amounts to about $10^{-8}$. Note that $x = k\,r$ in the definition of the Green's function, eq.(\ref{eq:a1a2}). 

\subsection*{Expansion around $x=\infty$}
The numerical values of the terms obtained in the asymptotic development in powers of $1/x$ have been slightly modified as to minimize the error in the interval $x=[2 ,\infty]$, such that the error is smaller than $10^{-8}$ everywhere, sacrificing the convergence rate for large values of $x$, where the $K_0$ and $K_1$ are almost zero anyway.
\[
	K_0(x) = e^{-x}\sqrt{\frac{\pi}{2x}}\Bigl(1 - 0.12498635 x^{-1} + 0.06988090 x^{-2} -  0.06781674 x^{-3}
\]
\[
	+ 0.07504864 x^{-4}  - 0.06422396 x^{-5} + 0.027170459 x^{-6} \Bigr).
\]

\[
	K_1(x) = e^{-x}\sqrt{\frac{\pi}{2x}} \Bigl(1 + 0.3749837 x^{-1}   - 0.11667051 x^{-2} + 0.09601858 x^{-3}  
\]
\[
	- 0.09962106 x^{-4} + 0.08313676 x^{-5}  -0.03484904 x^{-6} \Bigr).
\]

\subsection*{Expansion around $x=0$}
With $t = \log(x/2) + \gamma_E$, where $\gamma_E$ is the Euler-Mascheroni constant: 
\[
	K_0(x) = -t + \frac{x^2}{4}\Bigl( 1-t +\frac{x^2}{32} \Bigl( 3-2 t +\frac{x^2}{108} \Bigl( 11-6 t + \frac{x^2}{128} \Bigl( 25-12 t 
\]
\[
+ \frac{x^2}{500} \Bigl( 137-60 t + \frac{x^2}{48} \Bigl( 49-20 t + \frac{x^2}{1372} \Bigl( 363-149 t \Bigr) \Bigr) \Bigr) \Bigr) \Bigr) \Bigr) \Bigr),
\]

\[
	K_1(x)= \frac{1}{x} + \frac{x}{4} \Bigl( -1+2 t +\frac{x^2}{16} \Bigl( -5+4 t + \frac{x^2}{18} \Bigl( -5+3 t +\frac{x^2}{384} \Bigl( -47+24 t 
\]
\[
	+\frac{x^2}{200} \Bigl( -131+60 t +\frac{x^2}{60} \Bigl( -71+30 t + \frac{x^2}{784} \Bigl( -353+140 t  \Bigr) \Bigr) \Bigr) \Bigr) \Bigr) \Bigr) \Bigr).
\]

One shall note that the inner expansion of $K_0$ and $K_1$ bring the singular behavior to the Greens function of the Brinkman equation. $A_1$ inherits a leading order weak singularity $-\log(x/2)$ from $K_0$, which may be subtracted from the series and integrated analytically on singular boundary elements. 
The leading order singularity of $K_1/x$ is $1/x^2$ which can be right away cancelled out against $1/x^2$ in the eq.(\ref{eq:a1a2}) for $A_1$ and $A_2$. The strongly singular behavior in the Green functions stress tensor $\mathbf{T}$ stems from the last term, $K_1(x)\,x-2A_2$.


\begin{thebibliography}{}

	\bibitem{Whitesides}
	{George M. Whitesides},
	{The origins and the future of microfluidics},
	\textit{Nature} \textbf{442}, 368-373 (2006),
	{doi:10.1038/nature05058}.
	
	\bibitem{seemann12}
	{R. Seemann, M. Brinkmann, T. Pfohl and S. Herminghaus},
	{Droplet Based Microfluidics},
	\textit{Reports on Progress in Physics} \textbf{75} (1), 016601 (2012).
	{doi:10.1088/0034-4885/75/1/016601}.

	\bibitem{Borhan}
	{N. R. Gupta, A. Nadim, H. Haj-Hariri and A. Borhan}, 
	{A Numerical Study of the Effect of Insoluble Surfactants on the Stability of a Viscous Drop Translating in a {Hele-Shaw} Cell},
	\textit{J. of Colloid and Interface Science}
	\textbf{252} (1), 236–248 (2002).
	{doi:10.1006/jcis.2002.8441}.

		\bibitem{BoosThess}
	{W. Boos, A. and Thess},
	{Thermo-capillary flow in a {Hele-Shaw} cell},
	\textit{J. Fluid Mech.}
	\textbf{352} 305-330	(1997),
	{doi:10.1017/S0022112097007477}.

		 \bibitem{bush97}
	 {John W. M. Bush},
	 {The anomalous wake accompanying bubbles rising in a thin gap: a mechanically forced Marangoni flow},
	 \textit{Journal of Fluid Mechanics} \textbf{352}, 283-303 (1997),
	 {doi:10.1017/S0022112097007350}.

	\bibitem{Bazhlekov}
	{I. B. Bazhlekov, P. D. Anderson and Han E. H. Meijer},
	{Nonsingular boundary integral method for deformable drops in viscous flows},
	\textit{Physics of Fluids} \textbf{16}, 1064 (2004),
	{doi:10.1063/1.1648639}.
	
	\bibitem{Wrobel}
	{L. C. Wrobel, D. Soares Jr., and C. L. D. Bhaumik},
	{Drop deformation in Stokes flow through converging channels},
	\textit{Engineering Analysis with Boundary Elements} \textbf{33} (7), 993–1000 (2009),
	{doi:10.1016/j.enganabound.2009.01.009}.
	
	\bibitem{Nadim}
	{A. Nadim, A. Borhan and H. Haj-Hariri},
	{Tangential Stress and Marangoni Effects at a Fluid–Fluid Interface in a Hele–Shaw Cell},
	\textit{Journal of Colloid and Interface Science} \textbf{181}, 159–164 (1996).
	{doi:10.1006/jcis.1996.0367}.

	\bibitem{brinkman}
	{H. C. Brinkman},
	{A Calculation of the Viscous Force Exerted by a Flowing Fluid on a Dense Swarm of Particles.}
	\textit{Applied Scientific Research}
	\textbf{1} (1), 27–34 (1949).
	
	\bibitem{deville14}
	{U. E. Langlois, M. O. Deville},
	{Slow Viscous Flow, 2nd edition},
	{p. 149-151},
	\textit{Springer} {2014},
	{doi:10.1007/978-3-319-03835-3}.


	\bibitem{Gallaire}
	{F. Gallaire, P. Meliga, P. Laure and C. N. Baroud.}
	{Marangoni-induced force on a drop in a Hele-Shaw cell}
	\textit{Physics of Fluids}
	\textbf{26} (6), 062105, (2013).
	{doi:10.1063/1.4878095}

	\bibitem{park84}
	{C.-W. Park and G. M. Homsy},  
	{Two-phase displacement in Hele Shaw cells: theory.}
	\textit{J. Fluid Mech.}
	\textbf{139} 291-308 (1984).
	 {doi:10.1017/S0022112084000367}.

	\bibitem{zaleski}
  	{ G. Tryggvason, R. Tryggvason, and S. Zaleski},
  	{Direct Numerical Simulations of Gas-Liquid Multiphase Flows},
  	\textit{Cambridge University Press}, p. 170 (2011).
  	{doi:10.1017/CBO9780511975264}.

	\bibitem{amberg10}
	{A. Carlson, M. Do-Quang, and G. Amberg},
	{Droplet dynamics in a bifurcating channel},
	\textit{International Journal of Multiphase Flow} \textbf{36} (5),397-405 (2010).
	{doi:10.1016/j.ijmultiphaseflow.2010.01.002.}


	
	\bibitem{pozrikidis}
	{C. Pozrikidis},
	{A Practical Guide to Boundary Element Methods with the Software Library BEMLIB.},
	\textit{Chapman and Hall/CRC} May 2002, 161 -189.
	{ISBN 978-1-58488-323-4}.
	
	\bibitem{wrobel}
	{Luiz C. Wrobel, Delfim Soares Jr. and Claire L. Das Bhaumik},
	{Drop deformation in Stokes flow through converging channels},
	\textit{Engineering Analysis with Boundary Elements} \textbf{33} (7), 993-1000 (2009).
	{doi:10.1016/j.enganabound.2009.01.009.}
	


	\bibitem{Power}
	{G. Zhu, A. A. Mammoli and H. Power},
	{A 3-D indirect boundary element method for bounded creeping flow of drops},
	\textit{Engineering Analysis with Boundary Elements} \textbf{30} (10), 856–868 (2006),
	{doi:10.1016/j.enganabound.2006.07.002}.

	\bibitem{shelley93}
	{W.-S. Dai, M.J. Shelley},
	{A numerical study of the effect of surface tension and noise on an expanding Hele-Shaw bubble},
	\textit{Physics of Fluids}
	\textbf{5} (9), 2131-2146 (1993).
	{http://dx.doi.org/10.1063/1.858553}.
	
	\bibitem{shelley94}
	{T. Hou, J. S. Lowengrub and M. J. Shelley},
	{Removing the Stiffness from Interfacial Flows with Surface Tension},		
	\textit{JCP} \textbf{114}, 312-338 (1994).
	{doi:10.1006/jcph.1994.1170}.

	\bibitem{lapack}
	{LAPACK Users' Guide} 1999,
	\textit{Society for Industrial and Applied Mathematics},
	{http://www.netlib.org/lapack}.

	\bibitem{openmp}
  	{OpenMP Architecture Review Board},
  	{{http://www.openmp.org/mp-documents/spec30.pdf}}, 
  	\textit{OpenMP application program interface version 3.0} (2008).		
	
	\bibitem{nagela}
	{M. Nagel and F. Gallaire},
	{A new prediction of wavelength selection in radial viscous fingering involving normal and tangential stresses},
	\textit{Physics of Fluids}
	\textbf{25, no. 12} 124107 (2013).
	{doi:10.1063/1.4849495}
	
		\bibitem{cordero14}
	{Camilo Ulloa, Alberto Ahumada, and Maria Luisa Cordero, },
	{Effect of confinement on the deformation of microfluidic drops},
	\textit{Phys. Rev. E} \textbf{89} (3), {033004} (2014).
	{doi:10.1103/PhysRevE.89.033004}.

	\bibitem{barets}
	{Q. Brosseau, J. Vrignon and J.-C. Baret},
	{Microfluidic Dynamic Interfacial Tensiometry},		
	\textit{Soft Matter} \textbf{10} (17), {3066-3076} (2014). 
	{doi:10.1039/c3sm52543k}.
	
	\bibitem{cabral}
	{J. Cabral and S. D. Hudson},
	{Microfluidic approach for rapid multicomponent interfacial tensiometry},
	\textit{Lab on a Chip} \textbf{6} 427-436 (2006).
	{doi:10.1039/B511976F}.	

	\bibitem{nagel2}
 	{P.-T. Brun, M. Nagel and F. Gallaire},
	{Generic path for droplet relaxation in microfluidic channels},
  	\textit{Phys. Rev. E}
	\textbf{88} (4), 043009 (2013),
	{doi:10.1103/PhysRevE.88.043009}.

	\bibitem{nagel3}
	{M. Nagel, P.-T. Brun and F. Gallaire},
   	{A numerical study of droplet trapping in microfluidic devices},
   	\textit{Physics of Fluids}
	\textbf{26} (3), 032002 (2014),
	{doi:10.1063/1.4867251}.
	

\end{thebibliography}
\end{document}